\documentclass[aps,prd,twocolumn,eqsecnum,nofootinbib]{revtex4}
\usepackage{amssymb}
\usepackage{amsmath}
\usepackage{amsfonts}
\usepackage{longtable}
\newcommand{\be}{\begin{equation}}
\newcommand{\ee}{\end{equation}}
\newcommand{\bea}{\begin{eqnarray}}
\newcommand{\eea}{\end{eqnarray}}
\newcommand{\bes}{\begin{subequations}}
\newcommand{\ees}{\end{subequations}}
\newcommand{\nn}{\nonumber}
\newcommand{\ve}{\varepsilon}
\usepackage{bm}
\usepackage{graphicx}

\begin{document}

\title{Analysis of spin precession in binary black hole systems including quadrupole-monopole interaction}
\author{\'{E}tienne Racine}
\affiliation{Maryland Center for Fundamental Physics, Department of Physics, University of Maryland, College Park, MD 20742}
\date{\today}

%%%%%%%%%%%%%%%%%%%%%%%%%%%%%%%%%%%%%%%%%%%%%%%%
\begin{abstract}

We analyze in detail the spin precession equations in binary black hole systems, when the tidal torque on a Kerr black hole due to quadrupole-monopole coupling is taken into account. We show that completing the precession equations with this term reveals the existence of a conserved quantity at 2PN order when averaging over orbital motion. This quantity allows one to solve the (orbit-averaged) precession equations exactly in the case of equal masses and arbitrary spins, neglecting radiation reaction. For unequal masses, an exact solution does not exist in closed form, but we are still able to derive accurate approximate analytic solutions. We also show how to incorporate radiation reaction effects into our analytic solutions adiabatically, and compare the results to solutions obtained numerically. For various configurations of the binary, the relative difference in the accumulated orbital phase computed using our analytic solutions versus a full numerical solution vary from $\sim 0.3\%$ to $\sim 1.8\%$ over $\sim 80 - 140$ orbital cycles accumulated while sweeping over the orbital frequency range $\sim 20 - 300\, {\rm Hz}$. This typically corresponds to a discrepancy of order $\sim 5-6$ radians. While this may not be accurate enough for implementation in LIGO template banks, we still believe that our new solutions are potentially quite useful for comparing numerical relativity simulations of spinning binary black hole systems with post-Newtonian theory.  They can also be used to gain more understanding of precession effects, with potential application to the gravitational recoil problem, and to provide semi-analytical templates for spinning, precessing binaries.

\end{abstract}
%%%%%%%%%%%%%%%%%%%%%%%%%%%%%%%%%%%%%%%%%%%%%%%%
\maketitle

\section{Introduction and Summary}

Coalescing black hole binary systems are a class of sources of gravitational radiation that can potentially be detected by observatories such as LIGO \cite{LIGO}, VIRGO \cite{VIRGO}, GEO \cite{GEO} and TAMA \cite{TAMA}. Data analysis algorithms searching for such sources are based on the method of matched filtering, which essentially consists of correlating the detector output with a bank of (theoretical) template signals. For detection to be at all possible, these templates must be accurate to $\sim$ 1 radian over hundreds to thousands of cycles. In the regime where the black holes are separated by a large enough distance, the post-Newtonian (PN) expansion is expected to be accurate enough to be useful in generating gravitational waveform templates, provided one carries the expansion to high enough (3.5 PN) order (e.g. see \cite{BlanchetLiving} and references therein). 

While the distribution of spins for black holes in binaries is still a matter of ongoing research, it is generally thought very plausible that these spins may be large. It is therefore important to include in template banks waveforms that account for black hole spins, as they leave significant imprints on the signal, e.g. strong modulation of the waveform amplitude and phase \cite{Apostolatos, BCV2, Pan}. Ignoring effects of black hole spins could very well lead to missing such binaries when analyzing interferometer data. 

In order to predict the contribution of the spins to the gravitational waveform to the accuracy required by the data analysis process, the time evolution of the spins must be computed at least up to 2.5 post-Newtonian order (e.g. see \cite{Kidder, BBF} and references therein). However in this paper we restrict attention, for simplicity, to the precession equations accurate at 2PN order, i.e. including leading order spin-orbit and spin-spin terms. To that order the spin evolution equations, which will also be loosely referred to as "precession equations" throughout, take the functional form

\bea
\frac{d\bm{S}_{1,2}}{dt} &=&  f_0(t) \bm{J} \times \bm{S}_{1,2} + f_1(t) \bm{S}_{2,1} \times \bm{S}_{1,2} \nn \\
&& + \,\,{\rm terms \,\, of \,\,higher \,\,PN \,\, order}, \label{precessfunc}
\eea
where $\bm{S}_{1,2}$ are the spin 3-vectors, $\bm{J}$ is the system's total angular momentum, and  $f_{0,1}(t)$ are known, implicit functions of coordinate time. In the literature the functions $f_{0,1}$ are generally computed assuming a model of spinning point-particles, i.e. the precession equations are derived from solving the PN expanded Einstein equations assuming a distributional stress-energy tensor that contains mass monopole and spin dipole degrees of freedom. However this procedure ignores the precession due to the coupling of the mass quadrupole moment of each spinning hole to the tidal field of the companion\footnote{This is essentially the general relativistic version of the well-known phenomenon of the precession of the equinoxes on Earth.}. Since the mass quadrupole moment $Q$ of a Kerr black hole scales in order of magnitude as $Q \sim S^2/M$, where $M$ is here the black hole mass and $S$ the magnitude of its angular momentum, the precession of the hole's spin due its mass quadrupole contributes at the same PN order as the spin-spin term in Eq.(\ref{precessfunc}), but is nevertheless often omitted in the literature, with the following notable exceptions. 

Let us first point out the works of Barker and O'Connell \cite{barker} and Damour \cite{EOB}, who consider quadrupole-monopole coupling in the precession problem. Barker and O'Connell derive generic precession equations from a Hamiltonian point of view, while Damour focuses to the problem of spinning binary black holes, which allows a very compact Hamiltonian formulation at 2PN order. Next we note the contributions of Poisson \cite{Poisson} and Pan {\it et al.} \cite{Pan}, who have looked at some contributions of this term in the context of gravitational wave data analysis. In Ref.\cite{Poisson} Poisson argues that since the quadrupole-monopole precession contributes at the same PN order as spin-spin precession, the main features of the work of Apostolatos {\it et al.}, who study extensively spin precession in binary black holes at leading order in post-Newtonian theory, should remain qualitatively unchanged by the addition of the quadrupole-monopole precession term, as it is a sub-leading term. In Ref.\cite{Pan}, Pan {\it et al.} devote a subsection to a comparison of gravitational wave templates with and without the quadrupole-monopole precession, their conclusion being that the contribution of the quadrupole-monopole precession can be neglected. Lastly it is worth mentioning the works of Gergely, Keresztes and Mik\'{o}czi \cite{g1,g2}, who consider the quadrupole-monopole contributions to both conservative and dissipative pieces of orbital dynamics of compact binaries, and in particular provide a complete solution to the binary's radial motion.

However in light of the analyses of Poisson\cite{Poisson} and Pan {\it et al.}\cite{Pan}, it is understandable that a complete study of the precession equations including the quadrupole-monopole contribution for black hole binaries has not yet been completed. The main purpose of this paper is to perform this analysis carefully, being motivated by the following two considerations. The first element is consistency: in black hole binaries the quadrupole-monopole term contributes at the same PN order than spin-spin coupling. Thus if spin-spin coupling is taken into account in constructing template waveforms, then so should quadrupole-monopole coupling in principle. The second motivating element is essentially academic interest in an unsolved problem, coupled with the fact that sometimes order of magnitude estimates miss important facts and more careful investigations reveal hidden features in a problem. In the analysis we present below, such a feature will be unravelled, namely the existence of a new conserved quantity in the orbit-averaged precession equations, when the quadrupole-monopole precession is taken into account.   

Our analysis is organized as follows. We first review in section \ref{QMint} the derivation of the spin precession term due to the mass quadrupole of a Kerr black hole. In section \ref{ConsQ} we show that including the mass quadrupole term in the orbit-averaged precession equations leads to the discovery of a new conserved quantity\footnote{This quantity is strictly conserved by the orbit-averaged precession equations if radiation reaction effects are neglected. If radiation damping is taken into account, it evolves over the radiation reaction timescale.}. We then show in section \ref{AnSols} that this conserved quantity allows one to solve the orbit-averaged precession equations exactly in the case of equal masses and arbitrary spins, provided one neglects radiation reaction. In the case of unequal masses, we derive an accurate perturbative analytic solution. These solutions allow one to identify all the relevant frequencies characterizing the precession and nutation motions of each spin relative to the total angular momentum axis. Finally in section \ref{RR}, we derive approximate analytic solutions to the precession equations taking into account radiation damping, and compare these solutions to numerically integrated solutions.

\section{Quadrupole-monopole interaction}\label{QMint}

In Newtonian gravity, it is well-known that a body possessing a mass quadrupole moment (or any higher order mass multipole moment) placed into a prescribed gravitational tidal field (e.g created by a companion object) experiences a torque, which leads to a non-trivial time evolution of its spin. This effect is, for example, what causes the precession of the Earth's spin axis (precession of the equinoxes). A simple computation of the torque $\tau_i$ in Newtonian gravity, using a coordinate system $(t,\bm{x})$ that is mass-centered on the body,  yields

\bea\nn
\tau_i &=&  \int d^3x \, \rho(\bm{x},t) \epsilon_{ijk} x_j [-\partial_k \Phi_{\rm ext}(\bm{x},t)] \nn \\
&=& \int d^3x \, \, \rho(\bm{x},t) \epsilon_{ijk} x_j \nn \\
&& \times \partial_k \sum_{l=0}^\infty \frac{1}{l!}G_{<i_1i_2 ... i_l>}(t)x^{<i_1}x^{i_2}... x^{i_l>}, \label{torque}
\eea
where the brackets $<...>$ denote the symmetric trace-free projection, $\rho(\bm{x},t)$ is the mass density of the body, $\Phi_{\rm ext}(\bm{x},t)$ is the Newtonian potential created by an external source, and where\footnote{Note that the $l=0$ term does not contribute in (\ref{torque}) due to the spatial derivative $\partial_k$, and thus we do not need defining $G_{<i_1 i_2 ... i_l>}(t)$ for $l=0$ in (\ref{GLdef}).}

\be\label{GLdef}
G_{<i_1i_2...i_l>}(t) =  -\partial_{i_1}\partial_{i_2}...\partial_{i_l} \Phi_{\rm ext}(\bm{x},t) \Big|_{\bm{x} = 0}.
\ee
The right-hand side of (\ref{GLdef}) is clearly symmetric, and Laplace's equation ensures that is also trace-free on all indices. Using the Newtonian definition of mass multipole moments

\be
M_{<i_1i_2...i_l>}(t) = \int d^3x \, \rho(\bm{x},t) \, x^{<i_1}x^{i_2}...x^{i_l>},
\ee
we obtain 

\bea
\tau_i &=& \epsilon_{ijk} \sum_{l=0}^\infty \frac{1}{l!}G_{k i_1...i_l} \int d^3x \, \rho(\bm{x},t) x^j x^{<i_1}x^{i_2}... x^{i_l>} \nn \\
&=& \epsilon_{ijk} \sum_{l=0}^\infty \frac{1}{l!}G_{k i_1...i_l} \int d^3x \, \rho(\bm{x},t) x^{<j} x^{i_1}x^{i_2}... x^{i_l>} \nn \\
&=& \epsilon_{ijk} \sum_{l=0}^\infty \frac{1}{l!} M_{ji_1... i_l}G_{ki_1... i_l},\label{precess}
\eea
where the second line follows from the fact that $x^j x^{<i_1}x^{i_2}... x^{i_l>}$ and  $x^{<j} x^{i_1}x^{i_2}... x^{i_l>}$ differ by trace terms, which do not contribute since $\epsilon_{ijk}G_{ki_1... j ... i_l} = 0$. The form (\ref{precess}) of the torque follows the syntax of Damour, Soffel and Xu \cite{DSX}. When specializing to a binary system and restricting attention solely to the quadrupole-monopole interaction, we find 

\be\label{QMprecess}
\frac{dS_1^i}{dt} = 3\epsilon_{ijk} Q_1^{<jm>} M_2 \frac{n^{<k}n^{m>}}{r^3},
\ee
where $Q_1^{<jm>}$ is the mass quadrupole of body 1, $M_2$ is the mass of body 2, $n^i$ is a unit vector pointing from body 1 to body 2 and $r$ is orbital separation. While Eq.(\ref{precess}) has been revisited here in the context of Newtonian gravity, its validity in the case of a binary black hole system in the regime where the post-Newtonian expansion is valid has been proven rigorously by means of a surface-integral method by Racine in \cite{Racine}. For that situation, the mass multipole moments of each object are defined through their imprints in the far-field metric. Thus the mass quadrupole moment entering Eq.(\ref{QMprecess}) in the case of a spinning black hole is well-known and is given by (e.g. see Poisson \cite{Poisson})

\be\label{QKerr}
Q_1^{<ij>} = -\frac{1}{M_1}\left(S_1^i S_1^j - \frac{1}{3}\delta^{ij}S_1^kS_1^k\right),
\ee
where $S_1^i = \chi_1 M_1^2 \hat{S}_1^i$, $\chi_1$ being the dimensionless magnitude of the black hole spin, and $\hat{S}_1^i$ the unit vector along the spin of body 1. Substituting (\ref{QKerr}) into (\ref{QMprecess}) we obtain

\be
\frac{d\bm{S}_1}{dt} = \frac{3}{r^3}\frac{M_2}{M_1} (\hat{\bm{n}} \cdot \bm{S}_1) \hat{\bm{n}} \times \bm{S}_1
\ee
for the quadrupole-monopole precession term. When adding this term to the precession equations typically found in the literature (see e.g. \cite{Kidder,ThorneHartle}), the full precession equations are 

\bes\label{Sevol}
\bea
\frac{d\bm{S}_1}{dt} &=& \frac{1}{r^3}\left\{\left[2 + \frac{3}{2q}\right]\bm{L}_N - \bm{S}_2 
%\right. \nn \\ && \left. 
+ \frac{3\mu}{M_1}[\hat{\bm{n}}\cdot\bm{S}_0]\hat{\bm{n}}\right\} \times \bm{S}_1, \nn \\ \, \\ 
\frac{d\bm{S}_2}{dt} &=& \frac{1}{r^3}\left\{\left[2 + \frac{3q}{2}\right]\bm{L}_N - \bm{S}_1 
%\right. \nn \\ && \left. 
+ \frac{3\mu}{M_2}[\hat{\bm{n}}\cdot\bm{S}_0]\hat{\bm{n}}\right\} \times \bm{S}_2, \nn \\ 
\eea
\ees
where $q = M_1/M_2$, $\mu = M_1M_2/(M_1+M_2)$, and where 

\be\label{S0Damourdef}
\bm{S}_0 = \left(1 + \frac{M_2}{M_1}\right)\bm{S}_1 +  \left(1 + \frac{M_1}{M_2}\right)\bm{S}_2.
\ee
The vector $\bm{S}_0$ was originally introduced by Damour in the 2PN Hamiltonian for spinning binary black holes \cite{EOB}. To close the system (neglecting radiation-reaction for the time being), one uses conservation of total angular momentum\footnote{Strictly speaking, the true conserved angular momentum at 2PN is expanded as $\bm{J} = \bm{L} + \bm{S} =  \bm{L}_N + \bm{L}_{PN} + \bm{L}_{SO} + \bm{L}_{2PN} + \bm{S}$ (see e.g. \cite{Kidder}). However we can neglect all post-Newtonian corrections to $\bm{L}$, i.e. we take $\bm{L} = \bm{L}_N$, since these corrections introduce terms in the precession equations which are of higher PN order than the terms considered in this paper.} $\bm{J} = \bm{L}_N + \bm{S}_1 + \bm{S}_2$ to derive the precession equation for the (Newtonian) orbital angular $\bm{L}_N$, which is

\be\label{LNevol}
\frac{d\bm{L}_N}{dt} = \frac{1}{r^3}\left[\left(\frac{1}{2}\bm{S} + \frac{3}{2}\bm{S}_0\right) \times \bm{L}_N + \frac{3\mu}{M}(\hat{\bm{n}}\cdot\bm{S}_0)\hat{\bm{n}}\times\bm{S}_0\right],
\ee
where $\bm{S} = \bm{S}_1 + \bm{S}_2$. One may also recognize the often-encountered combination

\be
2\bm{S}_{\rm eff} = \frac{1}{2}\bm{S} + \frac{3}{2}\bm{S}_0.
\ee

\section{A new conserved quantity}\label{ConsQ}

From Eqs.(\ref{Sevol}), one can estimate the timescale for precession of the spins $\tau_{\rm p}$ to be  $\tau_{\rm p} \sim r^3/|\bm{L}_N| \sim (r/\mu) \tau_{\rm orb}$, $\tau_{\rm orb}$ being the orbital period. Thus it is customary to average the precession equations (\ref{Sevol})-(\ref{LNevol}) over an orbital period if one is interested in the secular evolution of the spins, i.e. neglecting small fluctuations occurring over the orbital timescale. This average over orbital motion is performed using (see e.g. Schnittman \cite{JS}) 

\bes
\bea
\langle \frac{1}{r^3}\rangle &=& \frac{1}{a^3(1-e^2)^{3/2}}, \\
\langle \frac{n^i n^j}{r^3} \rangle &=& \frac{1}{2a^3(1-e^2)^{3/2}}\big(\delta^{ij} - \hat{\bm{L}}_N^i\hat{\bm{L}}_N^j\big),
\eea
\ees
where $a$ is the semi-major axis and $e$ is the eccentricity. Defining $d \equiv a\sqrt{1-e^2}$, we then obtain the following system of evolution equations

\bes\label{Sdot}
\bea
\frac{d\bm{S}_1}{dt} &=& \frac{1}{2d^3}\left\{\left[4 + \frac{3M_2}{M_1} - \frac{3\mu}{M_1}\lambda\right]\bm{L}_N + \bm{S}_2\right\}\times\bm{S}_1, \label{S1dot} \nn \\
&& \, \\
\frac{d\bm{S}_2}{dt} &=& \frac{1}{2d^3}\left\{\left[4 + \frac{3M_1}{M_2} - \frac{3\mu}{M_2}\lambda\right]\bm{L}_N + \bm{S}_1\right\}\times\bm{S}_2, \label{S2dot}\nn \\
&& \, \\
\frac{d\bm{L}_N}{dt} &=& \frac{1}{2d^3}\left\{\bm{S} + 3\left[1 + \frac{\mu}{M}\lambda\right]\bm{S}_0\right\}\times \bm{L}_N, \label{Ldot}
\eea
\ees
where

\be\label{lambda}
\lambda = \frac{\bm{L}_N\cdot\bm{S}_0}{|\bm{L}_N|^2}.
\ee
A key property of system (\ref{Sdot}) is that it possesses a previously unknown conserved quantity, whose identification was only possible by completing the precession equations with the quadrupole-monopole interaction. This conserved quantity turns to be $\lambda$, defined in Eq.(\ref{lambda}). This can be shown by a direct computation. First, clearly $|\bm{L}_N|^2$ is conserved from (\ref{Ldot}), giving

\bea
\frac{d\lambda}{dt} &\propto& \left\{\bm{L}_N \cdot \frac{d\bm{S}_0}{dt} + \bm{S}_0\cdot \frac{d\bm{L}_N}{dt}\right\} \nn \\
&\propto& \Big\{M_2 \bm{L}_N \cdot (\bm{S}_2 \times \bm{S}_1) + M_1 \bm{L}_N \cdot (\bm{S}_1 \times \bm{S}_2) \nn \\
&& + \mu \bm{S}_0 \cdot (\bm{S} \times \bm{L}_N) \Big\} \nn \\
&=&\Big\{M_2 \bm{L}_N \cdot (\bm{S}_2 \times \bm{S}_1) + M_1 \bm{L}_N \cdot (\bm{S}_1 \times \bm{S}_2) \nn \\
&& + (M_2\bm{S}_1 + M_1 \bm{S}_2) \cdot [(\bm{S}_1 + \bm{S}_2) \times \bm{L}_N)] \Big\} \nn \\
&=& \Big\{M_2 [\bm{L}_N \cdot (\bm{S}_2 \times \bm{S}_1) + \bm{S}_1 \cdot(\bm{S}_2 \times \bm{L}_N)] \nn \\
&& + M_1 [\bm{L}_N \cdot (\bm{S}_1 \times \bm{S}_2) + \bm{S}_2 \cdot (\bm{S}_1 \times \bm{L}_N)]  \Big\} = 0. \nn \\\label{proof}
\eea
It is important to note here that the conservation of $\lambda$ is a property of the orbit-averaged precession equations only. Repeating the same computation using (\ref{Sevol}) and (\ref{LNevol}) instead of (\ref{Sdot}) yields $d\lambda / dt \neq 0$. This implies, for example, that the conservation of $\lambda$ (when averaging over orbital motion) cannot be deduced directly from the 2PN Hamiltonian of spinning binary black holes \cite{EOB}, unless the orbital degrees of freedom are averaged out.

In addition, proof (\ref{proof}) is only valid over the precession timescale. Over the radiation reaction timescale, the orbital angular momentum evolution equation must be supplemented by the following dissipative term [restricting henceforth attention to orbits with negligible eccentricy, i.e. $d \rightarrow a + O(e^2)$]

\bea
\big[\dot{\bm{L}}_N\big]_{\rm rr} &=& - \frac{32}{5}\frac{\mu M^2}{d^4} \bm{L}_N .
\eea
The corresponding evolution equation for $\lambda$ is then easily shown to be

\be\label{lambdadot}
\frac{d\lambda}{dt} = \frac{32}{5}\frac{\mu M^2}{d^4} \lambda \equiv \gamma \lambda,
\ee
whose solution can formally be written as the following integral over orbital frequency $\omega = \sqrt{Md^{-3}}$

\be\label{lambdasol}
\lambda = \lambda_0 \exp \left[\int_{\omega_0}^\omega \frac{\gamma(\omega)}{\dot{\omega}} d\omega \right],
\ee
where $\lambda_0$ and $\omega_0$ are initial values.

\section{Analytic solutions in absence of radiation reaction}\label{AnSols}

The first step in solving the precession evolution equations is making use of the definition of total angular momentum $\bm{J} = \bm{L}_N + \bm{S}_1 + \bm{S}_2$ to eliminate $\bm{L}_N$ from (\ref{S1dot}) and (\ref{S2dot}). We then have

\bes\label{SdotII}
\bea
\frac{d\bm{S}_1}{dt} &=& \frac{1}{2d^3}\left\{\left[4 + \frac{3M_2}{M_1} - \frac{3\mu}{M_1}\lambda\right]\bm{J} \right. \nn \\ 
&& \left. - \left[\frac{3M}{M_1} - \frac{3\mu}{M_1}\lambda\right]\bm{S}_2\right\}\times\bm{S}_1 \label{S1dotII}, \\
\frac{d\bm{S}_2}{dt} &=& \frac{1}{2d^3}\left\{\left[4 + \frac{3M_1}{M_2} - \frac{3\mu}{M_2}\lambda\right]\bm{J} \right. \nn \\ 
&& \left. - \left[\frac{3M}{M_2} - \frac{3\mu}{M_2}\lambda\right]\bm{S}_1\right\}\times\bm{S}_2 \label{S2dotII},
\eea
\ees
where $\bm{J}$ is a constant vector. The usefulness of the conserved quantity $\lambda$ is now quite clear: it shows that the coefficients in the coupled evolution system for $\bm{S}_1$ and $\bm{S}_2$ are all constants. 

In the remainder of this section we solve (\ref{SdotII}) for both equal mass and unequal mass cases and arbitrary spins, generalizing the analytical analysis of Apostolatos {\it et al.} \cite{Apostolatos}, valid for equal masses, or when one of the spins is dynamically negligible.

%\footnote{More precisely they evolve only over the radiation reaction timescale, which is longer than the precession timescale by a factor $\sim (c/v_{\rm orb})^3$. We may therefore develop adiabatic-type solutions by treating these coefficients as constants when solving system (\ref{SdotII}) in the presence of radiation reaction, which is the subject of section \ref{RR}.}.  

\subsection{Equal mass case}

When $M_1 = M_2$, the structure of system (\ref{Sdot}) simplifies greatly, allowing in fact an exact solution. Defining the phase variable $\psi$ as\footnote{We keep the integral over $t^\prime$ explicit here, in anticipation of our treatment of radiation reaction in the next section.}

\be\label{psidef}
\psi = \int_0^t  \frac{1}{2d^3}\left[7 - \frac{3}{2}\lambda\right]|\bm{J}| dt^\prime,
\ee
and the parameter $\alpha$ as 

\be\label{alphadef}
\alpha = \frac{3}{|\bm{J}|}\left[ \frac{4 - \lambda}{14 - 3\lambda}\right],
\ee
the evolution of the spins is governed by

\bes
\bea
\frac{d\bm{S}_1}{d\psi} &=& \hat{\bm{J}} \times \bm{S}_1 - \alpha \bm{S}_2 \times \bm{S}_1, \\
\frac{d\bm{S}_2}{d\psi} &=& \hat{\bm{J}} \times \bm{S}_2 - \alpha \bm{S}_1 \times \bm{S}_2.
\eea
\ees
By adding these two equations together, we see immediately that the total spin $\bm{S} = \bm{S}_1 + \bm{S}_2$ simply precesses about $\hat{\bm{J}}$, i.e.

\be\label{Stotdot}
\frac{d\bm{S}}{d\psi} = \hat{\bm{J}} \times \bm{S}.
\ee
The solution to differential equation (\ref{Stotdot}) is

\be\label{Ssol}
\bm{S} = \bm{S}^\parallel_0 + \bm{S}^\perp_0 \cos \psi + (\hat{\bm{J}} \times \bm{S}^\perp_0) \sin \psi,
\ee
where $\bm{S}^\parallel_0$ and $\bm{S}^\perp_0$ are the projections of the initial total spin $\bm{S}_0$  [not to be confused with the combination of spins defined in Eq.(\ref{S0Damourdef}); henceforth $\bm{S}_0$ will always refer to initial total spin] in directions parallel and perpendicular to $\hat{\bm{J}}$ respectively. Equation (\ref{Ssol}) is the first part of the exact solution for the equal mass case. We again emphasize that this solution corresponds to simple precession of the total spin about the total angular momentum axis, with precession frequency $\Omega_{\rm p} = d\psi/dt$. 

Consider next the vector $\bm{\Delta} = \bm{S}_1 - \bm{S}_2$. Its evolution is governed by

\bea
\frac{d\bm{\Delta}}{d\psi} &=& \hat{\bm{J}} \times \bm{\Delta} - 2\alpha \bm{S}_2 \times \bm{S}_1 \nn \\
&=& \hat{\bm{J}} \times \bm{\Delta} - \alpha \bm{S} \times \bm{\Delta} \nn \\
&\equiv& \bm{\Omega} \times \bm{\Delta}. \label{Deltadot}
\eea
Taking one more derivative with respect to $\psi$ yields after some algebra

\be\label{Deltaddot}
\frac{d^2\bm{\Delta}}{d\psi^2} + (\hat{\bm{J}} - \alpha\bm{S})^2\bm{\Delta} = (\hat{\bm{J}} \cdot \bm{\Delta})(\hat{\bm{J}} - 2\alpha\bm{S}) + \alpha^2(\bm{S}\cdot\bm{\Delta})\bm{S}.
\ee
Now $\bm{S}^2$, $\hat{\bm{J}}\cdot\bm{S}$ and $\bm{S}\cdot\bm{\Delta}$ are all constants of motion, which are thus known in terms of input initial conditions. However $\hat{\bm{J}}\cdot\bm{\Delta}$ is not yet known. Hence before (\ref{Deltaddot}) can be solved explicitly, we must first solve for $\hat{\bm{J}}\cdot\bm{\Delta}$. This is done directly by contracting (\ref{Deltaddot}) with $\hat{\bm{J}}$, giving

\be
\frac{d^2 (\hat{\bm{J}}\cdot \bm{\Delta})}{d\psi^2} + \alpha^2\bm{S}^2 (\hat{\bm{J}}\cdot \bm{\Delta}) = \alpha^2 (\bm{S}\cdot\bm{\Delta})(\bm{S}\cdot\hat{\bm{J}}).
\ee
The solution is immediate

\be
(\hat{\bm{J}}\cdot\bm{\Delta}) = A \cos\sigma + B\sin\sigma+ (\hat{\bm{S}}\cdot\bm{\Delta})(\hat{\bm{S}}\cdot\hat{\bm{J}}),
\ee
where the phase variable $\sigma$ is defined below in Eq.(\ref{sigmadef}). To fix the integration constants $A$ and $B$, we use the initial conditions and Eq.(\ref{Deltadot}) to obtain

\bea
(\hat{\bm{J}}\cdot\bm{\Delta}) &=& \mu_{JS}(\hat{\bm{S}}_0\cdot\bm{\Delta}_0) + [(\hat{\bm{J}}\times \hat{\bm{S}}_0)\cdot(\bm{\Delta_0}\times \hat{\bm{S}}_0)]\cos\sigma \nn \\
&& + \hat{\bm{J}}\cdot(\bm{\Delta}_0 \times \hat{\bm{S}}_0) \sin\sigma, \label{JdotDelta}
\eea
where $\mu_{JS }= \hat{\bm{S}}_0\cdot \hat{\bm{J}}$. Equation (\ref{JdotDelta}) shows that the projection of the difference of the spins along the total angular momentum axis oscillates at a frequency slower than the total spin precession frequency, since $d\sigma / dt = (\alpha S) \Omega_{\rm p}$ [cf. Eq.(\ref{sigmadef}) below]. This implies that each individual spin $\bm{S}_{1,2}$ exhibits a nutation motion at frequency $\Omega_{\rm n} = (\alpha S) \Omega_{\rm p}$. [Here we define nutation motion as non-trivial time evolution of the projection of a given spin vector along $\hat{\bm{J}}$.] Since the nutation frequency is proportional to $\alpha$, it is entirely due to spin-spin coupling. Barker, Byrd and O'Connell \cite{BBOC} have computed spin nutation frequencies in generic binary systems. However it is difficult to compare their results to ours since they rely on a succession of approximations, along with a somewhat implicit, time-dependent definition of the nutation frequency. In addition it is not entirely clear how their definition of nutation motion maps onto ours. Since our result is exact, up to the same PN order as considered by \cite{BBOC}, and provides a clean, constant value for the nutation frequency along with a precise definition of nutation motion, we believe our result is an improvement over the work of  \cite{BBOC}. However our result is applicable only to binary systems of Kerr black holes.

Substituting (\ref{JdotDelta}) into (\ref{Deltaddot}), one can then solve (\ref{Deltaddot}) using the retarded Green's function for the simple harmonic oscillator. However it turns out to be more convenient to simply project (\ref{Deltaddot}) along the remaining two independent directions, which we choose to be along $\bm{S}_0^\perp$ and $\hat{\bm{J}}\times\bm{S}_0^\perp$, in order to match the form of solution (\ref{Ssol}) for the total spin. The remaining evolution equations are then

\begin{widetext}
\bea
\frac{d^2}{d\psi^2}(\bm{S}_0^\perp \cdot \bm{\Delta}) &=& -\bm{\Omega}^2 (\bm{S}_0^\perp \cdot \bm{\Delta}) + \alpha^2(\bm{S}_0^\perp)^2(\bm{S}_0 \cdot \bm{\Delta}_0)\cos \psi - 2\alpha(\bm{S}_0^\perp)^2(\hat{\bm{J}}\cdot\bm{\Delta})\cos \psi \nn \\
&=& -\bm{\Omega}^2 (\bm{S}_0^\perp \cdot \bm{\Delta}) + \alpha (\bm{S}_0^\perp)^2(\hat{\bm{S}}_0 \cdot \bm{\Delta}_0)(\alpha S - 2\mu_{JS})\cos \psi \nn \\
&& - \alpha(\bm{S}_0^\perp)^2 \Big\{\big[(\hat{\bm{J}}\times \hat{\bm{S}}_0)\cdot(\bm{\Delta_0}\times \hat{\bm{S}}_0)\big]\big[\cos\big[(1+\alpha S)\psi\big] + \cos\big[(1-\alpha S)\psi\big]\big]\Big\} \nn \\
&& + \alpha(\bm{S}_0^\perp)^2\Big\{\big[\hat{\bm{J}}\cdot(\hat{\bm{S}}_0 \times \bm{\Delta}_0)\big]\big[\sin\big[(1+\alpha S)\psi\big] - \sin\big[(1-\alpha S)\psi\big]\big]\Big\} ,
\eea
\bea 
\frac{d^2}{d\psi^2} [(\hat{\bm{J}} \times \bm{S}_0^\perp) \cdot \bm{\Delta}]&=& -\bm{\Omega}^2  [(\hat{\bm{J}} \times \bm{S}_0^\perp) \cdot \bm{\Delta} ]+ \alpha^2(\bm{S}_0^\perp)^2(\bm{S}_0 \cdot \bm{\Delta}_0)\sin \psi - 2\alpha(\bm{S}_0^\perp)^2(\hat{\bm{J}}\cdot\bm{\Delta})\sin \psi \nn \\
&=& -\bm{\Omega}^2 [(\hat{\bm{J}} \times \bm{S}_0^\perp) \cdot \bm{\Delta}] + \alpha (\bm{S}_0^\perp)^2(\hat{\bm{S}}_0 \cdot \bm{\Delta}_0)(\alpha S - 2\mu_{JS})\sin \psi \nn \\
&& - \alpha(\bm{S}_0^\perp)^2 \Big\{\big[(\hat{\bm{J}}\times \hat{\bm{S}}_0)\cdot(\bm{\Delta_0}\times \hat{\bm{S}}_0)\big]\big[\sin\big[(1+\alpha S)\psi\big] + \sin\big[(1-\alpha S)\psi\big]\big]\Big\} \nn \\
&& - \alpha(\bm{S}_0^\perp)^2\Big\{\big[\hat{\bm{J}}\cdot(\hat{\bm{S}}_0 \times \bm{\Delta}_0)\big]\big[\cos\big[(1+\alpha S)\psi\big] - \cos\big[(1-\alpha S)\psi\big]\big]\Big\}. 
\eea
The solutions to these forced harmonic oscillators are 

\bea
\bm{S}_0^\perp \cdot \bm{\Delta} &=& [\bm{S}_0^\perp \cdot \bm{\Delta}_0]\cos\varpi+  [\bm{S}_0^\perp \cdot (\hat{\bm{\Omega}}_0 \times \bm{\Delta}_0)]\sin\varpi + S(1-\mu_{JS}^2) \Bigg\{ (\hat{\bm{S}}_0 \cdot \bm{\Delta}_0)(\cos \psi - \cos \varpi) \nn \\
&& + \frac{1}{2}\big[(\hat{\bm{J}}\times \hat{\bm{S}}_0)\cdot(\bm{\Delta_0}\times \hat{\bm{S}}_0)\big]\left[\frac{\cos\varpi - \cos(\psi - \sigma)}{(1-\mu_{JS})} - \frac{\cos\varpi - \cos(\psi + \sigma)}{(1+\mu_{JS})}\right] \nn \\ 
&& + \frac{1}{2} \big[\hat{\bm{J}}\cdot(\hat{\bm{S}}_0 \times \bm{\Delta}_0)\big]\left[\frac{\delta_+\sin \varpi - \sin(\psi + \sigma)}{(1+\mu_{JS})} + \frac{\delta_-\sin \varpi - \sin(\psi - \sigma)}{(1-\mu_{JS})}\right]  \Bigg\},\label {SperpdotDelta}
\eea
\bea
(\hat{\bm{J}} \times \bm{S}_0^\perp) \cdot \bm{\Delta}&=& [(\hat{\bm{J}} \times \bm{S}_0^\perp) \cdot \bm{\Delta}_0]\cos\varpi +  [(\hat{\bm{J}} \times \bm{S}_0^\perp) \cdot (\hat{\bm{\Omega}}_0 \times \bm{\Delta}_0)]\sin\varpi + S(1-\mu_{JS}^2)\Bigg\{ (\hat{\bm{S}}_0 \cdot \bm{\Delta}_0)(\sin \psi - \Omega^{-1}\sin \varpi)\nn \\
&& + \frac{1}{2}\big[(\hat{\bm{J}}\times \hat{\bm{S}}_0)\cdot(\bm{\Delta_0}\times \hat{\bm{S}}_0)\big]\left[\frac{\delta_-\sin \varpi - \sin(\psi - \sigma)}{(1-\mu_{JS})}-\frac{\delta_+\sin\varpi - \sin(\psi + \sigma)}{(1+\mu_{JS})} \right] \nn \\ 
&& \frac{1}{2} \big[\hat{\bm{J}}\cdot(\hat{\bm{S}}_0 \times \bm{\Delta}_0)\big]\left[\frac{\cos(\psi+\sigma) - \cos\varpi}{(1+\mu_{JS})} + \frac{\cos(\psi-\sigma) - \cos\varpi}{(1-\mu_{JS})}\right] \Bigg\},\label{JcrossSperpdotDelta}
\eea
\end{widetext}
where 

\bea
\Omega &=& \sqrt{\bm{\Omega}^2} = (1 - 2\alpha S\mu_{JS} + \alpha^2 S^2)^{1/2}, \\
\delta_\pm &=& \Omega^{-1}(1 \pm \alpha S), 
\eea
and where the phase variables $\varpi$ and $\sigma$ are defined as

\bea
\varpi &=& \int_0^\psi \Omega \, d\psi^\prime, \label{varpidef}\\
\sigma & = & \int_0^\psi \alpha S \, d\psi^\prime. \label{sigmadef}
\eea
Here we note that the components of the difference of the spins perpendicular to the total angular momentum axis contain terms which oscillate at four different frequencies. One of these frequencies is again $\Omega_{\rm p}$, and the other three are close, but not equal to the total spin precession frequency $\Omega_{\rm p}$, namely $(1 - 2\alpha S\mu_{JS} + \alpha^2 S^2)^{1/2} \Omega_{\rm p}$ and $(1\pm \alpha S)\Omega_{\rm p}$. Spin-spin coupling is again responsible for the appearance of these three additional precession frequencies.

The complete vector $\bm{\Delta}$ is then given by combining (\ref{JdotDelta}),(\ref{SperpdotDelta}) and (\ref{JcrossSperpdotDelta}) to give 

\be\label{DeltaSol}
\bm{\Delta} = (\hat{\bm{J}}\cdot\bm{\Delta})\hat{\bm{J}} + \frac{(\bm{S}_0^\perp \cdot \bm{\Delta})}{(1-\mu_{JS}^2)S^2}\bm{S}_0^\perp + \frac{[(\hat{\bm{J}} \times \bm{S}_0^\perp) \cdot \bm{\Delta}]}{(1-\mu_{JS}^2)S^2}\hat{\bm{J}} \times \bm{S}_0^\perp.
\ee
The individual spins are recovered from $\bm{S}_{1,2} = \frac{1}{2}(\bm{S} \pm \bm{\Delta})$. We emphasize once more an important property of this solution: each individual spin exhibits nutation at frequency $\Omega_{\rm n} = (\alpha S) \Omega_{\rm p}$, which is typically much slower than the precession frequency of the total spin. Since the nutation frequency is directly proportional to $\alpha$, this clearly shows that this effect is entirely induced by spin-spin coupling. The evolution of the components of the spins transverse to $\hat{\bm{J}}$ is rather complicated, since it contains components oscillating at four different frequencies, which are $\Omega_{\rm p}$, $(1 - 2\alpha S\mu_{JS} + \alpha^2 S^2)^{1/2}\ \Omega_{\rm p}$ and $\Omega_{\rm p} \pm \Omega_{\rm n}$. Again if one neglects spin-spin coupling, all four frequencies of the precession motion reduce to a single precession frequency, namely $\Omega_{\rm p}$.

Note also that in the case where, say, $|\bm{S}_2| \ll |\bm{S}_1|$, i.e. the spin of body 2 is dynamically negligible, then $\bm{S}_1$ becomes the total spin $\bm{S}$ and clearly the vector $\bm{\Delta}$ also becomes equal to the total spin. Obviously solution (\ref{Ssol}) for the total spin is still valid in that situation. It is straightforward to check that by setting $\bm{\Delta}_0 = \bm{S}_0$ in (\ref{DeltaSol}), one recovers $\bm{\Delta} = \bm{S}$ for all time.
 
\subsection{General case}\label{sec:general}

Unlike the equal mass case, the general case of arbitrary spins and unequal masses unfortunately does not allow an exact solution in terms of elementary functions. We first prove this statement. The spin evolution equations in the general case are

\bes\label{SdotIII}
\bea
\frac{d\bm{S}_1}{d\psi} &=& \beta_1 \hat{\bm{J}}\times \bm{S}_1 - \alpha_1 \bm{S}_2 \times \bm{S}_1, \label{S1dotIII}\\
\frac{d\bm{S}_2}{d\psi} &=& \beta_2 \hat{\bm{J}}\times \bm{S}_2 - \alpha_2 \bm{S}_1 \times \bm{S}_2, \label{S2dotIII}
\eea
\ees
where

\bea
\alpha_{1,2} &=& \frac{1}{|\bm{J}|}\left[\frac{6}{M_{1,2}}\left(\frac{M - \mu\lambda}{14 - 3\lambda}\right)\right], \label{alpha12def} \\ 
\beta_{1,2} &=& \frac{4 + 3\frac{M_{2,1}}{M_{1,2}} -3\frac{\mu}{M_{1,2}}\lambda }{7 - \frac{3}{2}\lambda}. \label{beta12def}
\eea
Now consider the evolution equations satisfied by $\hat{\bm{J}}\cdot\bm{S}_{1,2}$ and $\bm{S}_1\cdot\bm{S}_2$. These are

\bea
\frac{d (\hat{\bm{J}}\cdot\bm{S}_{1,2})}{d\psi} &=& \pm \alpha_{1,2} \hat{\bm{J}}\cdot(\bm{S}_1 \times \bm{S}_2), \\
\frac{d (\bm{S}_1 \cdot \bm{S}_2)}{d\psi} &=& \beta_-\,\hat{\bm{J}}\cdot(\bm{S}_1 \times \bm{S}_2), \label{S1dotS2prime}
\eea
where $\beta_- = \beta_1 - \beta_2$. We see that all right-hand sides in the above equations are proportional to the same function, namely $\hat{\bm{J}}\cdot(\bm{S}_1 \times \bm{S}_2)$. Defining $\rho$ as

\be\label{rhodef}
\rho = \int_0^\psi \hat{\bm{J}}\cdot(\bm{S}_1 \times \bm{S}_2) d\psi^\prime,
\ee
we then have

\bea
\hat{\bm{J}}\cdot\bm{S}_{1,2} &=& \pm \alpha_{1,2} \, \rho + (\hat{\bm{J}}\cdot\bm{S}_{1,2})_0, \label{JdotS12rho}\\
\bm{S}_1 \cdot \bm{S}_2 &=& (\beta_-)\, \rho + (\bm{S}_1 \cdot \bm{S}_2)_0, \label{S1dotS2rho}
\eea
where the $0$ subscript denotes initial value. By taking a $\psi$ derivative of, say, Eq.(\ref{S1dotS2prime}) and using expressions (\ref{JdotS12rho})-(\ref{S1dotS2rho}), one can show that $\rho$ obeys the following differential equation

\bea
\frac{d^2\rho}{d\psi^2} &=& -3\beta_-\alpha_1\alpha_2\, \rho^2 - \Big[\beta_-^2 + 2\beta_-\,\hat{\bm{J}}\cdot\bm{\zeta}_0^- + (\bm{\zeta}_0^+)^2\Big]\rho  \nn \\
&& - \beta_-\Big[(\hat{\bm{J}}\times\bm{S}_1)\cdot(\hat{\bm{J}}\times\bm{S}_2) \Big]_0 \nn \\
&& + \bm{\zeta}_0^+ \cdot \Big[\hat{\bm{J}} \times (\bm{S}_1 \times \bm{S}_2)\Big]_0 , \label{d2rhodpsi2}
\eea
where $\bm{\zeta}^\pm_0 = (\alpha_2\bm{S}_1 \pm \alpha_1\bm{S}_2)_0$. Thus $\rho$ evolves as an anharmonic oscillator with a cubic term in its potential. In the equal mass case $\beta_- = 0$ and $\rho$ reduces to a simple harmonic oscillator, thus allowing a solution in terms of trigonometric functions. In the general case one can solve (\ref{d2rhodpsi2}) in terms of an elliptic integral\footnote{The author thanks Saul Teukolsky for pointing this out.}, more specifically in terms of the Weierstrass elliptic function (see e.g. \cite{WW}). However here, for sake of transparency and simplicity, we resort to a perturbative method to solve (\ref{d2rhodpsi2}), based on the existence of a small parameter that will be identified shortly. Equation (\ref{d2rhodpsi2}) is of the form

\be
\frac{d^2\rho}{d\psi^2} = - C_2 \rho^2 - C_1 \rho + C_0,
\ee
where $C_{0,1,2}$ are all constants, which can be read off directly from (\ref{d2rhodpsi2}). Defining $S_{1,2} = |\bm{S}_{1,2}|$ and the dimensionless variable $u = \rho/S_1S_2$, we have 

\bea
\frac{d^2u}{d\psi^2} &=& - S_1S_2C_2 u^2 - C_1 u + \frac{C_0}{S_1S_2} \nn \\
&\equiv& - B_2 u^2 - B_1 u + B_0. \label{d2udpsi2}
\eea 
Now the key element to realize is that $B_2$ is a small number, as it scales as 

\bea
B_2 &\sim& (\alpha_1 S_1)(\alpha_2 S_2)\sim \frac{M^2}{\bm{J}^2}\frac{S_1}{M_1} \frac{S_2}{M_2} \nn \\ 
&\sim& M_1M_2\frac{M^2}{\bm{J}^2} \chi_1 \chi_2 \nn \\
&\sim& \eta \frac{M^4}{\bm{J}^2} \chi_1\chi_2,
\eea
where (\ref{alpha12def}) has been used and where $\chi_{1,2} = S_{1,2}/M_{1,2}^2$ are dimensionless spins and $\eta = \mu/M$ is the symmetric mass ratio. 

Let us first consider the case where $\eta \rightarrow 0$, i.e. the extreme mass ratio limit. We adopt the convention here that $M_2 \rightarrow 0$, so $M_1 \rightarrow M$. In that case the total angular momentum is dominated by the total spin since $\bm{L}_N \sim \eta$. Therefore we have $J^2 \rightarrow M^4 \chi_1^2$ which gives 

\be
B_2 \rightarrow \eta \chi_2/ \chi_1 \rightarrow 0,
\ee
since $\chi_1$ is finite.

When masses are comparable in the regime where post-Newtonian theory is applicable, the total angular momentum is dominated by the orbital angular momentum, which yields $\bm{J}^2 \sim \eta^2 M^3 r$. This then yields 

\be
B_2 \sim \frac{\chi_1\chi_2}{\eta} \frac{M}{r} \sim \frac{\chi_1\chi_2}{\eta} v^2.
\ee
Since $\chi_1\chi_2 / \eta$ is typically a number of order unity when masses are comparable, $B_2$ scales as a 1PN correction term compared to the numbers $B_1$ and $B_0$, which are generically of order unity.

Provided that $u$ remains small enough, the anharmonic term in the equation of motion for $u$ generically remains a small correction to the harmonic driving term for all time. We shall assume this is the case in what follows, and verify {\it a posteriori} that this assumption indeed works quite well. We next define the variable 

\be
x = u - A_0,
\ee
where 

\be\label{A0def}
A_0 = \frac{-B_1 + \sqrt{B_1^2 + 4B_0B_2}}{2B_2}
\ee
is a constant chosen so that the evolution equation for $x$ takes the form

\be
\frac{d^2 x}{d\psi^2} = - \ve x^2 - w^2 x, \label{d2xdpsi2}
\ee   
where $\ve = B_2$ and 

\be
w^2 = B_1 + 2B_2A_0 = \sqrt{B_1^2 + 4B_0B_2}.
\ee
We now solve (\ref{d2xdpsi2}) by treating the anharmonic term as a small perturbation, with $\ve$ being our expansion parameter. Now it is well-known (see e.g. Goldstein \cite{Goldstein}) that the corrections to the frequency of an oscillator with a cubic perturbation of order $\ve$ in its potential scale as $\ve^2$. Thus it is necessary to solve (\ref{d2xdpsi2}) to order $\ve^2$ accuracy in order to capture this important effect. To that order the solution is

\begin{widetext}
\bea
x(\psi) &=& -\ve\left[\frac{x_0^2 + v_0^2}{2w^2} + \ve\frac{x_0^3}{3w^4}\right] + \left[x_0 + \ve\frac{x_0^2 + 2v_0^2}{3w^2} + \ve^2\frac{x_0(29x_0^2 - 55v_0^2)}{144w^4}\right]\cos(\Phi) \nn \\
&& + \left[v_0 - \ve\frac{2x_0v_0}{3w^2} + \ve^2\frac{5v_0(x_0^2 - 11v_0^2)}{144w^4}\right]\sin(\Phi) + \ve\left[\frac{x_0^2 - v_0^2}{6w^2} + \ve\frac{x_0(x_0^2 + 4v_0^2)}{9w^4}\right]\cos(2\Phi) \nn \\
&& + \ve\left[\frac{x_0v_0}{3w^2} + \ve\frac{v_0(-x_0^2 + 2v_0^2)}{9w^4}\right]\sin(2\Phi) + \ve^2\frac{x_0(x_0^2-3v_0^2)}{48w^4}\cos(3\Phi) + \ve^2\frac{v_0(3x_0^2 - v_0^2)}{48w^4}\sin(3\Phi), \label{xSol}
\eea
\end{widetext}
where $v_0 = (dx/d\Phi)_0$, with

\be\label{Phidef}
\Phi = \int_0^\psi w\sqrt{1 - \ve^2\frac{5[w^2x_0^2 + (x_0^\prime)^2]}{6w^6}}\, d\psi^\prime,
\ee
where $x^\prime_0 = (dx/d\psi)_0$. Since the components of each spin along $\hat{\bm{J}}$ are proportional to $x(\psi)$, one can identify the fundamental frequency associated with the nutation motion of each spin as $(d\Phi/d\psi)\Omega_{\rm p}$. However in the general case, the nutation motion contains higher harmonics of its fundamental, as opposed to the equal mass case where the nutation motion contains exactly one frequency. In terms of initial conditions of the spins, the remaining quantities appearing in (\ref{xSol}) are

\bea
x_0 &=& -A_0 = \frac{B_1 - \sqrt{B_1^2 + 4B_0B_2}}{2B_2}, \label{A0redef}
\eea
\bea
v_0 &=& \left[1 - \ve^2\frac{5[w^2x_0^2 + (x_0^\prime)^2]}{6w^6}\right]^{-1/2} \frac{x_0^\prime}{w},
\eea
\bea
x_0^\prime &=& \hat{\bm{J}}\cdot(\hat{\bm{S}}_1\times\hat{\bm{S}}_2)_0 ,
\eea
\bea
w &=& \left[B_1^2 + 4B_0B_2\right]^{1/4}, \label{wredef}
\eea
\bea
B_0 &=& \bm{\zeta}_0^+\cdot[\hat{\bm{J}}\times(\hat{\bm{S}}_1\times\hat{\bm{S}}_2)]_0 \nn \\
&&  - \beta_-[(\hat{\bm{J}}\times\hat{\bm{S}}_1)\cdot(\hat{\bm{J}}\times\hat{\bm{S}}_2)]_0, \label{B0redef}
\eea
\bea
B_1 &=& \beta_-^2 + 2\beta_-\,\hat{\bm{J}}\cdot\bm{\zeta}_0^- + (\bm{\zeta}_0^+)^2 ,
\eea
\bea
B_2 &=& \ve =   3(\beta_-)(\alpha_1S_1)(\alpha_2S_2). \label{B2redef}
\eea

Now recall that the function $x(\psi)$ [or equivalently $\rho(\psi)$] allows one to determine only three angles in the problem, namely the projections of $\bm{S}_{1,2}$ along the total angular momentum axis, and the angle between $\bm{S}_1$ and $\bm{S}_2$. To completely specify the unit vectors along $\bm{S}_1$ and $\bm{S}_2$, we require one more angle. One can arrive at a convenient choice for this angle by recalling that in the equal mass limit, the total spin simply precesses about the total angular momentum axis. So we choose the remaining angle to be the angle between the projection of $\bm{S}$ perpendicular to $\hat{\bm{J}}$ and the projection of $\bm{S}_0$ perpendicular to $\hat{\bm{J}}$. In the equal mass limit, this angle is simply given by $\psi$ [cf. Eq.(\ref{Ssol})]. We now determine an approximate formula for its equivalent in the general case. For unequal masses, the evolution equations for $\bm{S}$ and $\bm{\Delta}$ can be shown to be

\bes\label{dotperps}
\bea
\frac{d\bm{S}}{d\psi} &=& \frac{1}{2}\left[\beta_+\hat{\bm{J}}\times\bm{S} + \beta_-\, \hat{\bm{J}}\times\bm{\Delta} + \alpha_-\,\bm{\Delta}\times\bm{S}\right], \nn \\
&& \, \\
\frac{d\bm{\Delta}}{d\psi} &=& \frac{1}{2}\left[\beta_+\hat{\bm{J}}\times\bm{\Delta} + \beta_-\, \hat{\bm{J}}\times\bm{S} + \alpha_+\,\bm{\Delta}\times\bm{S}\right], \nn \\
\eea
\ees
where $\alpha_\pm = \alpha_1\pm\alpha_2$ and $\beta_\pm = \beta_1\pm\beta_2$. We next take another derivative of the evolution equation for the total spin, and project the result perpendicular to $\hat{\bm{J}}$ to obtain

\bes\label{dotperpsII}
\bea
\frac{d^2\bm{S}^\perp}{d\psi^2} &=& -\frac{1}{4}\Big[\beta_+^2 + \beta_-^2 + (2\beta_+\alpha_- + \alpha_+\beta_-)(\hat{\bm{J}}\cdot\bm{\Delta}) \nn \\
&&  - \alpha_-\beta_-(\hat{\bm{J}}\cdot\bm{S}) -\alpha_-\alpha_+(\bm{S}\cdot\bm{\Delta}) + \alpha_-^2\bm{\Delta}^2 \Big]\bm{S}^\perp \nn \\
&& - \frac{1}{4}\Big[2\beta_+\beta_- - (2\beta_+\alpha_- + \alpha_+\beta_-)(\hat{\bm{J}}\cdot\bm{S}) \nn \\
&& + \alpha_-\beta_-(\hat{\bm{J}}\cdot\bm{\Delta}) - \alpha_-^2(\bm{S}\cdot\bm{\Delta})+ \alpha_-\alpha_+\bm{S}^2\Big]\bm{\Delta}^\perp. \nn \\
\eea
\ees
The equivalent equation for $\bm{\Delta}^\perp$ is obtain by interchanging $\bm{S} \leftrightarrow \bm{\Delta}$ and $\alpha_\pm \leftrightarrow -\alpha_\mp$. The coefficients of $\bm{S}^\perp$ and $\bm{\Delta}^\perp$ depend on time only through the function $\rho(\psi)$, since $\hat{\bm{J}}\cdot\bm{S}_{1,2}$ and $\bm{S}_1\cdot\bm{S}_2$ are all directly proportional to $\rho$, which is an oscillatory function of $\psi$. Our key approximation is therefore to replace the coefficients of $\bm{S}^\perp$ and $\bm{\Delta}^\perp$ in (\ref{dotperpsII}) by their average\footnote{Since the evolution of $\rho$ is on the same timescale as the precession frequencies of $\bm{S}^\perp$ and $\bm{\Delta}^\perp$, one might object that this not a good approximation. However notice that the magnitude of the terms proportional to $\rho$ in (\ref{dotperpsII}) is small (of order $\sim \alpha S$) compared to the dominant constant terms. Thus we believe that this should be a reasonable approximation, at least to obtain some trial solutions which can then be compared against numerical integrations.}. The resulting system of equations for $\bm{S}^\perp$ and $\bm{\Delta}^\perp$ is then of the form

\bes\label{oscillators}
\bea
\frac{d^2\bm{S}^\perp}{d\psi^2} + \omega_S^2 \bm{S}^\perp = \kappa_S \bm{\Delta}^\perp, \\
\frac{d^2\bm{\Delta}^\perp}{d\psi^2} + \omega_\Delta^2 \bm{\Delta}^\perp = \kappa_\Delta \bm{S}^\perp,
\eea
\ees
where $\omega^2_{S,\Delta}$ and $\kappa_{S,\Delta}$ are constants given by

\bea
\omega_S^2 &=& \frac{1}{4}\Big[\beta_+^2 + \beta_-^2 + (2\alpha_-\beta_+ + \alpha_+\beta_-)\langle\hat{\bm{J}}\cdot\bm{\Delta}\rangle \nn \\
&& - \alpha_-\beta_-\langle\hat{\bm{J}}\cdot\bm{S}\rangle + \alpha_-^2\langle\bm{\Delta}^2\rangle - \alpha_-\alpha_+(S_1^2-S_2^2)\Big],\nn \\
\eea
\bea
\omega_\Delta^2 &=&  \frac{1}{4}\Big[\beta_+^2 + \beta_-^2 - (2\alpha_+\beta_+ + \alpha_-\beta_-)\langle\hat{\bm{J}}\cdot\bm{S}\rangle \nn \\
&& + \alpha_+\beta_-\langle\hat{\bm{J}}\cdot\bm{\Delta}\rangle + \alpha_+^2\langle\bm{S}^2\rangle - \alpha_-\alpha_+(S_1^2-S_2^2)\Big], \nn  \\
\eea
\bea
\kappa_S &=&  -\frac{1}{4}\Big[2\beta_+\beta_- - (2\alpha_-\beta_+ + \alpha_+\beta_-)\langle\hat{\bm{J}}\cdot\bm{S}\rangle \nn \\
&& + \alpha_-\beta_-\langle\hat{\bm{J}}\cdot\bm{\Delta}\rangle + \alpha_+\alpha_-\langle\bm{S}^2\rangle - \alpha_-^2(S_1^2-S_2^2) \Big], \nn \\
\eea
\bea
\kappa_\Delta &=& -\frac{1}{4}\Big[2\beta_+\beta_- + (2\alpha_+\beta_+ + \alpha_-\beta_-)\langle\hat{\bm{J}}\cdot\bm{\Delta}\rangle \nn \\
&& - \alpha_+\beta_-\langle\hat{\bm{J}}\cdot\bm{S}\rangle + \alpha_+\alpha_-\langle\bm{\Delta}^2\rangle - \alpha_+^2(S_1^2-S_2^2) \Big].\nn \\
\eea
The various averages appearing above are

\bea
\langle \hat{\bm{J}}\cdot\bm{S}\rangle &=& (\hat{\bm{J}}\cdot\bm{S})_0 + \alpha_-\langle \rho\rangle, \label{JSavg}\\
\langle \hat{\bm{J}}\cdot\bm{\Delta}\rangle &=& (\hat{\bm{J}}\cdot\bm{\Delta})_0 + \alpha_+\langle \rho\rangle, \label{JDavg}\\
\langle \bm{S}^2\rangle &=& \bm{S}_0^2 + 2\beta_-\langle\rho\rangle, \label{S2avg}\\
\langle \bm{\Delta}^2\rangle &=& \bm{\Delta}_0^2 - 2\beta_-\langle\rho\rangle, \label{D2avg} \\
\langle \rho \rangle &=& S_1S_2\left[A_0 - \ve\frac{x_0^2 + v_0^2}{2w^2} \right]. \label{rhoavg}
\eea
By postulating harmonic solutions, one finds the normal frequencies $\hat{\omega}_\pm$ of system (\ref{oscillators}) to be

\be
\hat{\omega}_\pm = \frac{1}{\sqrt{2}}\left[\omega_S^2+\omega_\Delta^2 \pm \sqrt{(\omega_S^2 - \omega_\Delta^2)^2 + 4\kappa_S\kappa_\Delta}\right]^{1/2}.
\ee
The general solutions for $\bm{S}^\perp$ and $\bm{\Delta}^\perp$ are thus linear combinations of harmonic functions oscillating at frequencies $\hat{\omega}_\pm$. In terms of initial conditions $\bm{S}^\perp_0, \bm{\Delta}^\perp_0, d{\bm{S}}^\perp_0/d\psi, d{\bm{\Delta}}^\perp_0/d\psi$, the solution for $\bm{S}^\perp$ is 

\bea
\bm{S}^\perp &=& \frac{1}{\hat{\omega}_+^2 - \hat{\omega}_-^2}\Bigg\{\Big[(\omega_S^2 - \hat{\omega}_-^2)\bm{S}^\perp_0 - \kappa_S\bm{\Delta}^\perp_0\Big] \cos(\Phi_+) \nn \\
&& -  \Big[(\omega_S^2 - \hat{\omega}_+^2)\bm{S}^\perp_0 - \kappa_S\bm{\Delta}^\perp_0\Big] \cos(\Phi_-) \nn \\
&& + \Big[(\omega_S^2 - \hat{\omega}_-^2)\frac{d\bm{S}^\perp_0}{d\psi} - \kappa_S\frac{d\bm{\Delta}^\perp_0}{d\psi}\Big] \frac{\sin(\Phi_+)}{\hat{\omega}_+}  \nn \\
&& -  \left[(\omega_S^2 - \hat{\omega}_+^2)\frac{d\bm{S}^\perp_0}{d\psi} - \kappa_S\frac{d\bm{\Delta}^\perp_0}{d\psi}\right] \frac{\sin(\Phi_-)}{\hat{\omega}_-}\Bigg\}, \nn \\ \label{SperpSol}
\eea
where the phases are

\be\label{Phipmdef}
\Phi_\pm = \int_0^\psi \hat{\omega}_\pm \, d\psi^\prime.
\ee
Then the angle $\Upsilon$ between $\bm{S}^\perp$ and $\bm{S}^\perp_0$ is obtained directly from (\ref{SperpSol})
as follows
\be
\tan \Upsilon = \frac{\bm{S}^\perp \cdot(\hat{\bm{J}}\times\bm{S}_0^\perp)}{\bm{S}^\perp \cdot \bm{S}_0^\perp}.
\ee
We have now assembled all the pieces necessary to write down final solutions for each spins. These are

\bea\label{S12sol}
\bm{S}_{1,2} &=&S_{1,2}\Big[(\cos\theta_{1,2})\hat{\bm{J}} + (\sin\theta_{1,2}\cos\varphi_{1,2})\hat{\bm{S}}^\perp_0 \nn \\ 
&& + (\sin\theta_{1,2}\sin\varphi_{1,2})\hat{\bm{J}}\times\hat{\bm{S}}^\perp_0 \Big].
\eea
The angles parametrizing the direction of unit vectors along each spin are the following

\bes\label{anglesfinal}
\bea
\theta_1 &=& {\rm arccos} \Big[(\alpha_{1}S_2)u + (\hat{\bm{J}}\cdot\hat{\bm{S}}_1)_0\Big], \\
\varphi_1 &=& \frac{1}{2}(\varphi_+  + \varphi_-), \\
\theta_2 &=& {\rm arccos}\Big[- (\alpha_{2}S_1)u + (\hat{\bm{J}}\cdot\hat{\bm{S}}_2)_0\Big],\\
\varphi_2 &=& \frac{1}{2}(\varphi_+ - \varphi_-), 
\eea
\ees
where
\bes
\bea
\varphi_+ &=& 2\,\,{\rm arctan} \left[\frac{D_1\tan\Upsilon - D_2}{D_1 + D_2\tan\Upsilon}\right],\\
\varphi_- &=& {\rm arctan}\left[\frac{-du/d\psi}{(\hat{\bm{S}}_1\cdot\hat{\bm{S}}_2)_0 + (\beta_-)u - \cos\theta_1\cos\theta_2}\right], \nn\\
\eea
\ees
with 

\bes
\bea
D_1 &=& (S_1\sin\theta_1 + S_2\sin\theta_2), \\
D_2 &=& (S_1\sin\theta_1 - S_2\sin\theta_2)\tan\left[\frac{\varphi_-}{2}\right].
\eea
\ees

Let us now look at some features of this solution. First of all, since the evolution of the components of the spins is described by non-linear oscillators (either with cubic potentials or time-dependent frequencies), the true frequency spectrum in the case of unequal masses contains an infinite number of components. However our approximate solutions demonstrate that only a few of these components are sufficient to obtain very good agreement with solutions obtained numerically, as depicted below in Fig.\ref{fig:angles}. More precisely we included here five different frequencies into our solutions for the spins, namely the fundamental nutation frequency $\Omega_p (d\Phi/d\psi)$, its first two harmonics $2\Omega_p (d\Phi/d\psi)$ and $3\Omega_p (d\Phi/d\psi)$, and the two frequencies appearing in $\bm{S}^\perp$ and $\bm{\Delta}^\perp$ which are $\hat{\omega}_\pm \Omega_p$.

To illustrate the accuracy of our approximate solution, we show below a plot of $\hat{\bm{S}}_1\cdot \hat{\bm{S}}_2$ and $\hat{\bm{S}}^\perp \cdot \hat{\bm{S}}^\perp_0$, the binary parameters being specified in the caption of Fig.\ref{fig:angles}. When specifying the initial binary configuration in the figure caption, the angles $\theta_{1,2}$ refer now to the angle between each initial spin and the initial orbital angular momentum, whereas the azimuthal angles $\varphi_{1,2}$ describe the projection of each initial spin in the initial orbital plane. They are not to be confused with the angles given in Eqs.(\ref{anglesfinal}), which use the total angular momentum as the $z$-axis. By convention we always pick $\varphi_1 = 0$ when specifying an initial binary configuration. We present  $\hat{\bm{S}}_1\cdot \hat{\bm{S}}_2 = \cos\theta_{12}$ and $\hat{\bm{S}}^\perp \cdot \hat{\bm{S}}^\perp_0$ in Fig.\ref{fig:angles} to provide contrast with the equal mass case. In the equal mass limit, $\cos\theta_{12}$ is a constant of motion (fluctuations over the orbital period are averaged out) and $\Upsilon$ increases proportional to time. Here clearly $\cos\theta_{12}$ here evolves with time. Note also that the evolution of $\cos\Upsilon$ contains very interesting structure, as it exhibits significant distortions from a pure sinusoidal shape. More specifically one can notice two turning points in the direction of precession of the component of the total spin perpendicular to total angular momentum. The first turning point occurs at $\sim 25$ orbits and the second at $\sim 30$ orbits. Our analytic solution captures this peculiar behavior rather well. We should point out however, that this is not a generic feature of the solutions. For most binary configurations we investigated, we did not notice turning points in the precession motion of the total spin.  

The agreement between our analytical prediction and numerical result for $\cos\theta_{12}$ is such that the numerical and analytical curves are indistinguishable on the figure. On this plot the maximum difference between the analytical approximation and the numerical result for $\cos\theta_{12}$ is $2.6 \times 10^{-5}$. This is evidence that our approximate solution to the cubic oscillator governing the evolution of $\cos\theta_{12}$ is very good. For $\cos\Upsilon$, one can distinguish the two curves by eye, but only barely. In this case, the maximum difference between the analytical approximation and the numerical result is $6.32 \times 10^{-2}$. This is not quite as accurate as $\cos\theta_{12}$, but recall that the angle $\Upsilon$ is obtain by approximating the actual evolution equations, rather than developing an approximate solution to the exact evolution equation. One could easily improve (\ref{SperpSol}) by taking into account the leading oscillating terms in $\rho$ in addition of considering its average value alone. This would generate terms oscillating at different frequencies than $\hat{\omega}_\pm$, and complicate solution (\ref{SperpSol}) considerably. For this first analytical investigation however, we believe the accuracy of (\ref{SperpSol}) as it stands is sufficient. 

\begin{figure}
\includegraphics[width=3.25in]{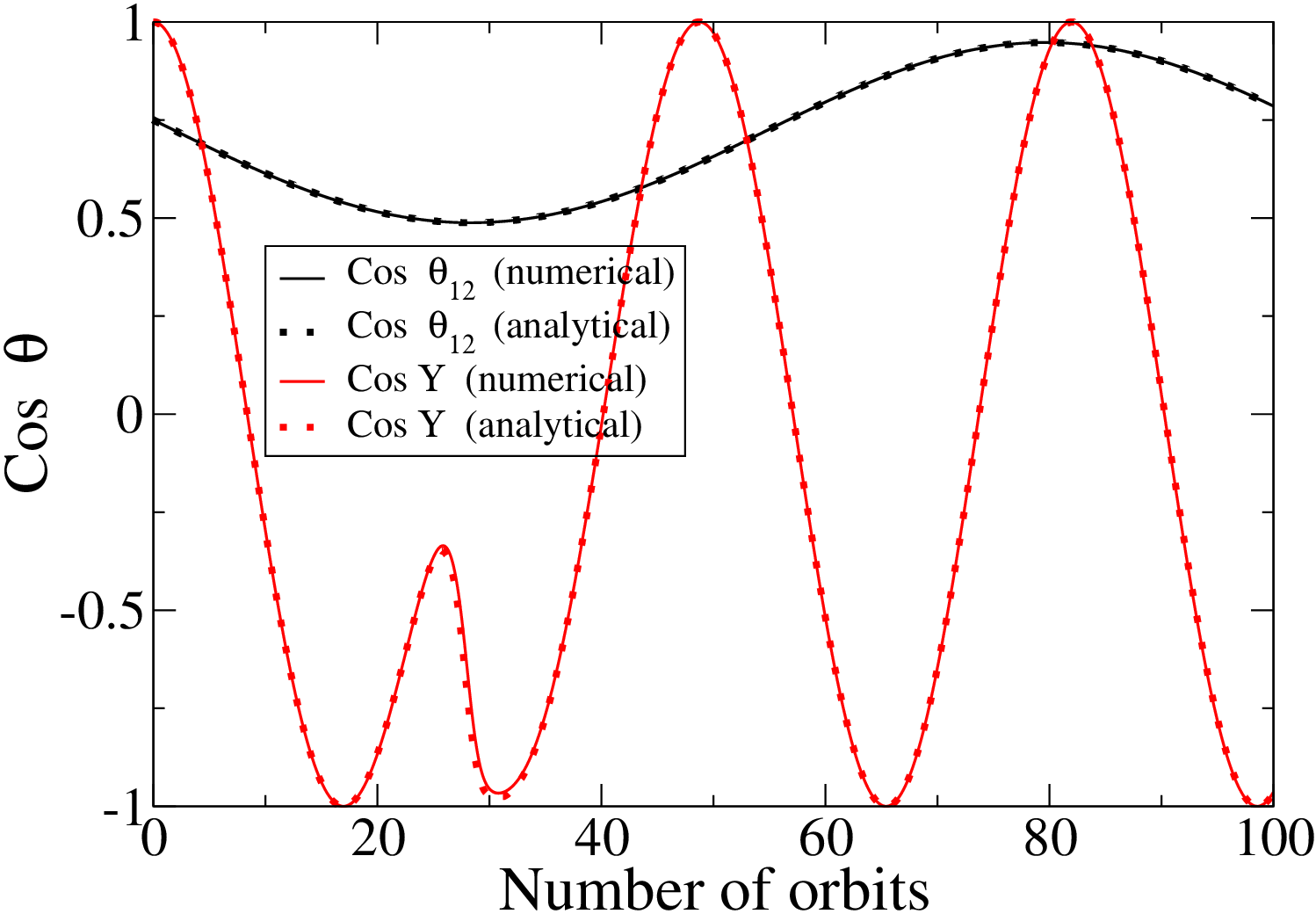}
\caption{\label{fig:angles} We plot, as function of time in units of orbital period, the angles $\cos\theta_{12} = \hat{\bm{S}}_1\cdot \hat{\bm{S}}_2$ and $\cos\Upsilon = \hat{\bm{S}}^\perp \cdot \hat{\bm{S}}^\perp_0$ for a binary with initial configuration $M = 10 M_\odot, f_{\rm orb} = 20\,{\rm Hz}, \chi_1 = \chi_2 = 0.8, \theta_1=\theta_2 = 30^\circ, \varphi_1 = 0^\circ$ and  $\varphi_2 = 90^\circ$. The mass ratio $q=M_1/M_2$ is here equal to 3/2.}
\end{figure}

In Fig.\ref{fig:angles}, we track the evolution of the system for 100 orbits at an orbital frequency of $20\, {\rm Hz}$, so that it lies near the low frequency limit of the LIGO band. For a $10\, M_\odot$ binary, this is clearly unrealistic, as the binary is expected to evolve significantly due to radiation reaction on that timescale. Thus incorporating radiation reaction in our solutions is relevant, and we address this issue in the next section.

\section{Effects of radiation reaction}\label{RR}

In the previous section we solved the precession equations analytically, but neglecting radiation reaction. This assumption implied that the total angular momentum $\bm{J}$ and the quantity $\lambda$ are constants of the motion, allowing derivation of (approximate) analytic solutions. In realistic applications, the precession equations need to be solved over the radiation reaction timescale, and it is therefore important to discuss the corrections to the solutions previously obtained due to radiation reaction. 

In the presence of radiation reaction, the total angular momentum evolves according to

\bea
\frac{d\bm{J}}{dt} &=& -\frac{32}{5}\frac{\mu M^2}{d^4}\bm{L}_N \nn \\
&\equiv& -\gamma (\bm{J}-\bm{S}). \label{dJdt}
\eea
In most astrophysical situations, the direction of $\bm{J}$ undergoes only a small precession around some average direction. This is known as simple precession \cite{Apostolatos}. In rare cases it may happen that the total spin nearly cancels the orbital angular momentum, which may lead to wild changes in the direction of $\bm{J}$. This is known as transitional precession \cite{Apostolatos}. In this paper we will restrict attention to simple precession alone and assume for simplicity that the direction of $\bm{J}$ remains constant throughout the inspiral.  

To illustrate our procedure to incorporate radiation-reaction into the solutions derived in the previous section, consider the precession of the total spin in the equal mass case. In absence of radiation reaction, recall the solution is given by

\be\label{Ssolnorr}
\bm{S} = [\hat{\bm{J}}\cdot\bm{S}_0]\hat{\bm{J}} + \bm{S}_0^\perp \cos\psi + [\hat{\bm{J}}\times\bm{S}_0^\perp]\sin\psi,
\ee
where

\be
\psi = \int_0^t \frac{1}{2d^3}\left[7 - \frac{3}{2}\lambda\right]|\bm{J}| dt^\prime.
\ee
When $\bm{J}$ depends on time, the evolution equation for $\bm{S}$ is still given by (\ref{Stotdot}). If the direction of $\hat{\bm{J}}$ were exactly constant, then the solution would still be (\ref{Ssolnorr}). However if $\hat{\bm{J}}$ varies, then corrections terms must be added to (\ref{Ssolnorr}). However in this paper, we shall implement radiation-reaction in the simplest possible way, which is given by the following prescription. One starts by writing down the spins for a given initial binary configuration using the solutions from section \ref{AnSols}. Then our solution which incorporates radiation-reaction is obtained by computing the phases $\psi, \varpi$ and $\sigma$ in the equal mass case, or $\Phi$ and $\Phi_\pm$ in the general case by performing explicitly the integrals appearing in Eqs.(\ref{psidef}),(\ref{varpidef}) and (\ref{sigmadef}), or (\ref{Phidef}) and (\ref{Phipmdef}) respectively. All the coefficients of all trigonometric functions appearing in the spins are assumed to be constants, determined by the initial conditions, i.e. we neglect possible evolution on the radiation reaction timescale of these coefficients. This approximation, along with the assumption that $\hat{\bm{J}}$ remains constant, are the main sources of errors in our solutions with radiation reaction included. However for a first analysis, we prefer to limit complications and leave improving the treatment of radiation-reaction beyond what is done here for future work. 

In order to calculate all required phases, the time evolution of the orbital frequency and of the magnitude of the total angular momentum are needed. We discuss these two topics next.

\subsection{Evolution of the orbital frequency}

We use the evolution equation for the orbital frequency accurate to 3.5 PN order, with spin effects included however only up to 2PN order (see e.g. \cite{Pan}), since this is the commonly used model employed in the data analysis community. It is given by 
\begin{widetext}
\bea
\frac{\dot{\omega}}{\omega^2} &=& \frac{96}{5}\eta(M\omega)^{5/3}\Bigg\{ 1 - \frac{743 - 924\eta}{336}(M\omega)^{2/3} - \left(\frac{1}{12}\sum_{i=1,2}\left[\chi_i \left(\hat{\bm{L}}_N\cdot\hat{\bm{S}}_i\right)\left(113\frac{M_i^2}{M^2} + 75\eta\right)\right] - 4\pi\right)(M\omega) \nn \\
&& + \left(\frac{34103}{18144} + \frac{13661}{2016}\eta + \frac{59}{18}\eta^2\right)(M\omega)^{4/3} - \frac{\eta}{48}\chi_1\chi_2\left[247(\hat{\bm{S}}_1\cdot\hat{\bm{S}}_2) - 721(\hat{\bm{L}}_N\cdot\hat{\bm{S}}_1)(\hat{\bm{L}}_N\cdot\hat{\bm{S}}_2)\right](M\omega)^{4/3} \nn \\
&& - \frac{\pi}{672}(4159 + 15876\eta)(M\omega)^{5/3} + \Bigg[\left(\frac{16447322263}{139708800} - \frac{1712}{105}\gamma_E + \frac{16}{3}\pi^2\right) - \left(\frac{273811877}{1088640} - \frac{451}{48}\pi^2 + \frac{91432}{13860} \right)\eta \nn \\
&& + \frac{541}{896}\eta^2 - \frac{5605}{2592}\eta^3 - \frac{856}{105}\log\left[16(M\omega)^{2/3}\right]\Bigg](M\omega)^2 - \left(\frac{4415}{4032} - \frac{358675}{6048}\eta - \frac{91495}{1512}\eta^2\right)\pi(M\omega)^{7/3}\Bigg\},\label{omega3.5PN}
\eea
\end{widetext}
where $\chi_{1,2} = |\bm{S}_{1,2}|/M_{1,2}^2$ are the dimensionless spins, $\eta = \mu/M$ is the symmetric mass ratio and $\gamma_E$ is the Euler-Mascheroni constant. It is worth mentioning here that there are other contributions to (\ref{omega3.5PN}) from the spins at 2PN order, namely the quadrupole-monopole interaction given by Poisson \cite{Poisson}, and radiative self-spin contributions computed by Mik\'{o}czi, Vas\'{u}th and Gergely \cite{Mikoczi}. There are also other contributions from spins at higher PN order which we also dropped here. Clearly it is formally inconsistent to neglect all these contributions, we nevertheless decided to do so for simplicity, and also to compare directly with a known model recently studied in the context of data analysis \cite{Pan}. This comparison should still be useful in estimating the errors generated by using our analytic solutions in (\ref{omega3.5PN}) instead of the solution provided by numerical integration.

Formulas for the total angular momentum and spin phases below involve the following function

\bea
I &=& \int_{\omega_0}^\omega \frac{\gamma(\omega)}{\dot{\omega}}d\omega \\
&=& \frac{32}{5}\frac{\eta}{M}  \int_{\omega_0}^\omega \frac{(M\omega)^{8/3}}{\dot{\omega}}\, d\omega. \label{Idef}
\eea
Substituting (\ref{omega3.5PN}) into (\ref{Idef}) and performing a Taylor expansion accurate to 3.5PN, one can in principle perform the integral, provided that some expression for the spins is provided. To simplify the problem, we use some sort of average values for the quantities $\hat{\bm{L}}_N\cdot\hat{\bm{S}}_{1,2}$ and $\hat{\bm{S}}\cdot\hat{\bm{S}}_{1,2}$ to avoid introducing complicated non-linearities in the problem. More precisely, we replace in (\ref{omega3.5PN}) $\hat{\bm{L}}_N\cdot\hat{\bm{S}}_{1,2}$ by $\langle\hat{\bm{J}}\cdot\hat{\bm{S}}_{1,2}\rangle$ and $\hat{\bm{S}}_1\cdot\hat{\bm{S}}_2$ by $\langle\hat{\bm{S}}_1\cdot\hat{\bm{S}}_2\rangle$, each average value being computed from Eqs.(\ref{JSavg})-(\ref{rhoavg}). Our final result for the function $I$ which appears throughout this section

\bea
I &=& \int_{y_0}^y \left[1 + \sum_{n=2}^7 c_n y^n\right] \frac{dy}{y} \nn \\
&=& \ln(y/y_0) + \sum_{n=2}^7 \frac{c_n}{n}\Big[y^n - y_0^n\Big], \label{Iresult}
\eea
where

\be
y = (M\omega)^{1/3}.
\ee
Writing (\ref{omega3.5PN}) as 

\be
\frac{\dot{\omega}}{\omega^2} = \frac{96}{5}\eta(M\omega)^{5/3}\left[1 + \sum_{n=2}^7 b_n y^n\right],
\ee
where the $b_n$'s are read off directly from right-hand side of (\ref{omega3.5PN}), the coefficients $c_n$ appearing in (\ref{Iresult}) are

\bes
\bea
c_2 &=& -b_2, \\
c_3 &=& -b_3, \\
c_4 &=& b_2^2 - b_4, \\
c_5 &=& 2b_2b_3 - b_5, \\
c_6 &=& -b_2^3 + b_3^2 + 2b_2b_4 - b_6, \\
c_7 &=& -3b_2^2b_3 + 2b_3b_4 + 2b_2b_5 - b_7.
\eea
\ees
Note that $b_6$, and thus $c_6$ actually depends logarithmically on frequency. Here we neglect this frequency dependence and use $b_6 \equiv b_6(t=0)$ instead, since the errors generated by this simplification are smaller that the errors coming from averaging the terms depending on the spins in $b_3$ and $b_4$. 

\subsection{Evolution of magnitude of total angular momentum}

We next move on to the computation of the magnitude $|\bm{J}|$ of the total angular momentum. This is required in order to compute the phases $\psi, \varpi, \sigma, \Phi$ and $\Phi_\pm$.

\subsubsection{Equal mass case}
In the equal mass case, the evolution equation for the total spin has the form

\be\label{dSdt}
\frac{d\bm{S}}{dt} \propto \bm{J} \times \bm{S},
\ee
From (\ref{dJdt}) and (\ref{dSdt}), we obtain directly

\be\label{dJdotSdt}
\frac{d(\bm{J}\cdot\bm{S})}{dt} = - \gamma\left[(\bm{J}\cdot\bm{S}) - \bm{S}^2\right].
\ee
Equation (\ref{dJdotSdt}) can be solved as

\be\label{JdotSsol}
\bm{J}\cdot\bm{S} = e^{-I}\left[(\bm{J}\cdot\bm{S})_0 - \bm{S}^2\right] + \bm{S}^2,
\ee
where $I$ is given by Eq.(\ref{Iresult}). Next contracting (\ref{dJdt}) with $\bm{J}$, one obtains

\be\label{dJ2dt}
\frac{d\bm{J}^2}{dt} = -2\gamma\left[\bm{J}^2 - \bm{J}\cdot\bm{S}\right],
\ee
whose solution is directly given by

\bea
\bm{J}^2 &=& e^{-2I}\left[\bm{J}_0^2 + 2\int_{\omega_0}^{\omega} e^{2I} \frac{\gamma}{\dot{\omega}}\bm{J}\cdot\bm{S} \,d\omega\right] \nn \\
&=& e^{-2I}\left[\bm{J}_0^2 + 2\left(e^I - 1\right)(\bm{J}\cdot\bm{S})_0 + \left(e^I - 1\right)^2 \bm{S}^2\right], \nn \\ \label{J2sol}
\eea
where (\ref{JdotSsol}) has been used to obtain the second line. Note incidentally that Eq.(\ref{J2sol}) is exact, and thus does not rely on the assumption of constant $\hat{\bm{J}}$ used throughout the paper. It is therefore also valid for spin configurations producing transitional precession. Since transitional precession occurs when $|\bm{J}| \rightarrow 0$, one can use (\ref{J2sol}) to predict if transitional precession can occur given an initial binary configuration (restricted for now to equal mass).

\subsubsection{General case}\label{sec:Jgen}

We now repeat the computation for the case of arbitrary masses and spins. Here however it is not possible to obtain $\bm{J}\cdot\bm{S}$ and hence $\bm{J}^2$ exactly, since the evolution equation for $\bm{J}\cdot\bm{S}$ contains a term proportional to $\bm{J}\cdot(\bm{S}_1\times\bm{S}_2)$, whose exact evolution is determined in terms of an elliptic integral. We start from

\bea
\frac{d(\bm{J}\cdot\bm{S})}{dt} &=& - \gamma\left[(\bm{J}\cdot\bm{S}) - \bm{S}^2\right] \nn \\
&& + \frac{d\psi}{dt}(\alpha_1 - \alpha_2)\bm{J}\cdot(\bm{S}_1\times\bm{S}_2) \nn \\ 
&=&  - \gamma\left[(\bm{J}\cdot\bm{S}) - \bm{S}^2\right] + |\bm{J}|(\alpha_1 - \alpha_2)\frac{d\rho}{dt} \nn \\ 
&=& - \gamma\left[(\bm{J}\cdot\bm{S}) - \bm{S}^2\right]  + \frac{3(M_2-M_1)}{7\mu}\frac{d\rho}{dt} \nn \\  \label{dJdotSdtgen}
\eea
where (\ref{alpha12def}) and (\ref{rhodef}) have been used, with terms of order $\lambda$ neglected in (\ref{alpha12def}). The solution to (\ref{dJdotSdtgen}) is 

\bea
\bm{J}\cdot\bm{S} &=& e^{-I}(\bm{J}\cdot\bm{S})_0  + e^{-I}\int_0^t e^{I} \gamma\bm{S}^2 dt^\prime \nn \\
&& + \frac{3(M_2-M_1)}{7\mu} e^{-I}\int_0^t e^{I} \frac{d\rho}{dt^\prime} dt^\prime. \label{JdotSgen}
\eea
Since $\rho$ is a function oscillating over the precession timescale, we will assume that the remaining integral in (\ref{JdotSgen}) averages out to a negligible contribution, and that we may substitute $\bm{S}^2$ by its initial average value $\langle \bm{S}^2\rangle_0$ [cf. Eqs.(\ref{S2avg}) and (\ref{rhoavg})]. The remainder of the calculation mirrors the manipulations of the previous subsection, which lead to

\bea
\bm{J}^2 &=& e^{-2I}\left[\bm{J}_0^2 + 2\left(e^I - 1\right)(\bm{J}\cdot\bm{S})_0 + \left(e^I - 1\right)^2 \langle\bm{S}^2\rangle_0\right]. \nn \\
\eea
This expression can be used to predict (roughly) the possible onset of transitional precession in the unequal mass case.

\subsection{Phases}

To complete the computation of radiation reaction effects, we need expressions for the phases $\psi, \sigma, \varpi, \Phi$ and $\Phi_\pm$ appearing in the solutions for the spins.

\subsubsection{Equal mass case}

The phases relevant to the equal mass solution are $\psi, \varpi$ and $\sigma$ [cf. Eqs.(\ref{psidef}), (\ref{varpidef}) and (\ref{sigmadef})]. We start with the phase $\psi$, which is given by

\bea
\psi &=& \int_0^t \frac{1}{2d^3}\left[7 - \frac{3}{2}\lambda\right]|\bm{J}|\, dt \nn \\
&=& \int_{\omega_0}^\omega \frac{\omega^2}{2M}\left[7 - \frac{3}{2}\lambda_0 e^I\right]e^{-I} \nn \\
&& \times \left[\bm{J}_0^2 + 2\left(e^I - 1\right)(\bm{J}\cdot\bm{S})_0 + (e^I - 1)^2\bm{S}^2\right]^{\frac{1}{2}} \frac{d\omega}{\dot{\omega}}, \nn \\
\eea
An exact evaluation of the above integral is impossible. It must therefore be evaluated numerically. We next turn our attention to the phase $\varpi$, which is given by

\bea
\varpi &=& \int_0^t \frac{\Omega}{2d^3}\left[7 - \frac{3}{2}\lambda\right]|\bm{J}|\, dt, \nn \\
&=& \int_{\omega_0}^\omega \Omega\frac{\omega^2}{2M}\left[7 - \frac{3}{2}\lambda_0 e^I\right]e^{-I} \nn \\
&& \times \left[\bm{J}_0^2 + 2\left(e^I - 1\right)(\bm{J}\cdot\bm{S})_0 + (e^I - 1)^2\bm{S}^2\right]^{\frac{1}{2}} \frac{d\omega}{\dot{\omega}}, \nn \\ \label{varpiofI}
\eea
where 

\be\label{OmegaofI}
\Omega = \left[1 - 2\alpha(\hat{\bm{J}}\cdot\bm{S}) + \alpha^2\bm{S}^2\right]^{1/2}.
\ee
Similarly to $\psi$, the phase $\varpi$ must be evaluated numerically. Lastly we look at the phase $\sigma$, given by

\bea
\sigma &=& \int_0^t \frac{\alpha S}{2d^3}\left[7 - \frac{3}{2}\lambda\right]|\bm{J}|\, dt  \nn \\
&=& \int_0^t \frac{3S}{4d^3}\left[4 - \lambda\right]\, dt \nn \\
&=& \int_{\omega_0}^\omega \frac{3S\omega^2}{4M}\left[4 - \lambda_0e^I\right] \frac{d\omega}{\dot{\omega}},
\eea
which we also evaluate numerically. To complete our analysis of the equal mass case, we give a plot of the precession frequencies as function of orbital frequency for the following initial binary configuration: $M = 10\, M_\odot$, $\chi_1 = \chi_2 = 0.5$, $\theta_1 = \theta_2 = 30^\circ$, $\phi_1 = 90^\circ$ and $\phi_2 = 0^\circ$. Above $M$ is the binary's total mass, $\chi_{1,2}$ are the dimensionless spins and the angles $\theta_{1,2}$ and $\phi_{1,2}$ are the polar and azimuthal angles giving the initial orientation of each spin in a coordinate system where the $z$ axis is given by the initial orbital angular momentum. The zero of the azimuthal coordinate $\phi$ is defined by the initial component of $\hat{\bm{S}}_1$ lying in the initial orbital plane.

\begin{figure}
\includegraphics[width=3.25in]{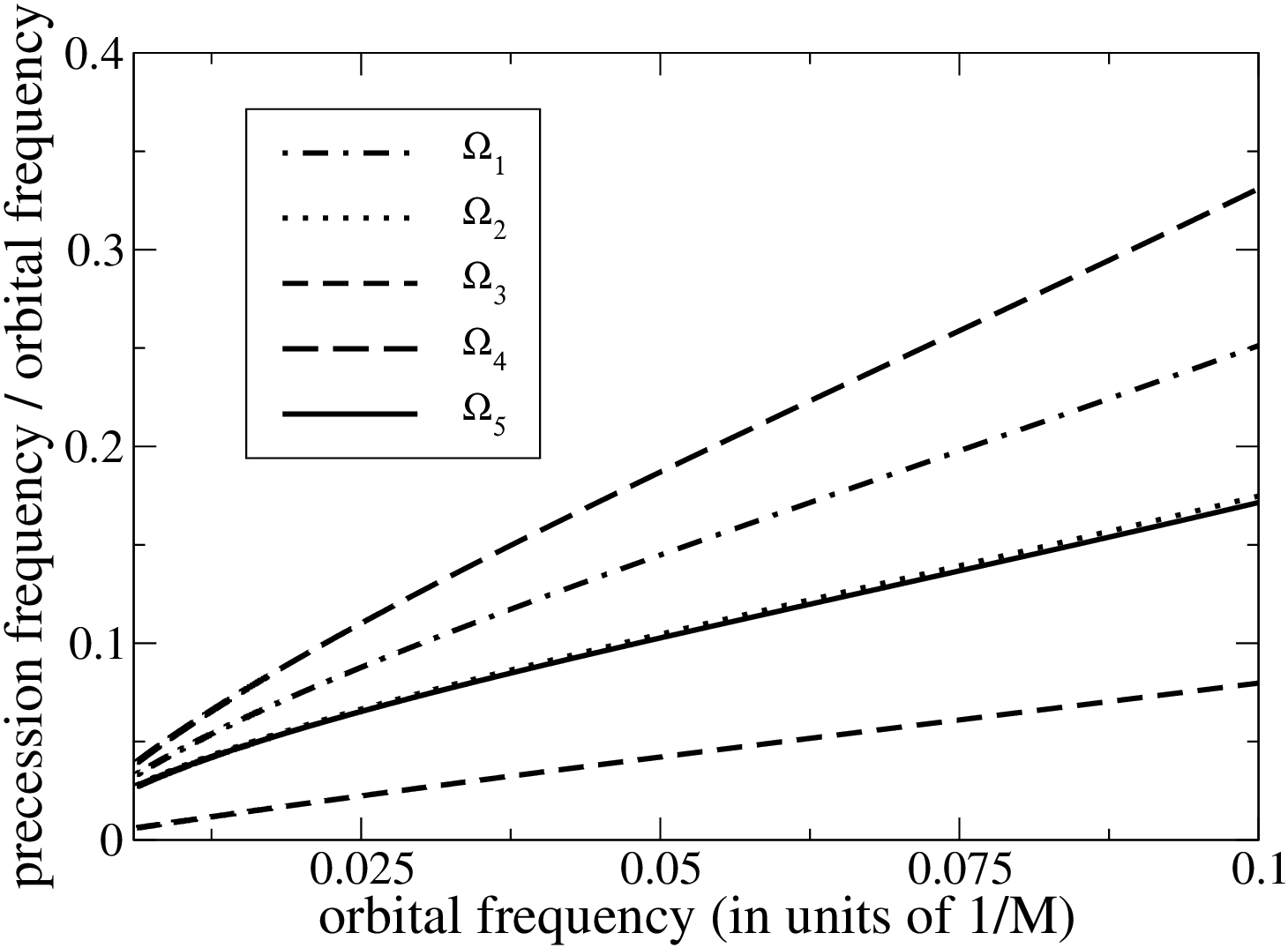}
\caption{\label{fig:freqs1} This is a plot of the precession frequencies appearing in the spins, for the equal mass case, as function of orbital frequency. The binary configuration at orbital frequency of 20 Hz is specified by the following parameters: $M = 10\, M_\odot$, $\chi_1 = \chi_2 = 0.8$, $\theta_1 = \theta_2 = 30^\circ$, $\phi_1 = 0^\circ$ and $\phi_2 = 90^\circ$. The precession frequencies are expressed in units of instantaneous orbital frequency. The frequencies shown in the legend are defined as $\Omega_1 = d\psi / dt$, $\Omega_2 = d\varpi / dt$, $\Omega_3 = d\sigma / dt$ , $\Omega_4 = \Omega_1 + \Omega_3$ and $\Omega_5 = \Omega_1 - \Omega_3$.}
\end{figure}
 
\subsubsection{General case}

The phases $\Phi$ and $\Phi_\pm$ appearing in the general case are defined in Eqs.(\ref{Phidef}) and (\ref{Phipmdef}) respectively. We first look at $\Phi$, which is

\bea
\Phi &=& \int_0^t \frac{w}{2d^3}\sqrt{1 - \ve^2\frac{5[w^2x_0^2 + (x_0^\prime)^2]}{6w^6}}\left[7 - \frac{3}{2}\lambda\right]|\bm{J}|dt^\prime. \nn \\
\eea
Similarly to the phase $\varpi$ [cf. Eq.(\ref{varpiofI})], since $w$ and $\ve$ are now functions of frequency, one can transform the above integral into the following

\bea
\Phi &=&  \int_{\omega_0}^\omega \frac{\omega^2}{2M}w\sqrt{1 - \ve^2\frac{5[w^2x_0^2 + (x_0^\prime)^2]}{6w^6}}\left[7e^{-I} - \frac{3}{2}\lambda_0\right] \nn \\
&& \times \left[\bm{J}_0^2 + 2\left(e^I - 1\right)(\bm{J}\cdot\bm{S})_0 + (e^I - 1)^2\bm{S}^2\right]^{\frac{1}{2}} \frac{d\omega}{\dot{\omega}}, \nn \\
\eea
which needs to be evaluated numerically. The time dependence of $w$ is obtained from (\ref{wredef}) by substituting the time-dependent quantities $\alpha_{1,2}(I)$ and $\beta_{1,2}(I)$ into (\ref{B0redef}) and (\ref{B2redef}). The last elements required to construct our approximate analytic solutions for the spins are the phases $\Phi_\pm$. They are given by

\bea
\Phi_\pm &=&  \int_{\omega_0}^\omega \frac{\omega^2}{2M}\left[7e^{-I} - \frac{3}{2}\lambda_0\right] \hat{\omega}_\pm \times \nn \\
&& \left[\bm{J}_0^2 + 2\left(e^I - 1\right)(\bm{J}\cdot\bm{S})_0 + (e^I - 1)^2\bm{S}^2\right]^{\frac{1}{2}} \frac{d\omega}{\dot{\omega}}, \nn \\
\eea
which cannot be evaluated in closed form. To complete our analysis of the general case, we give a plot of the precession frequencies as function of orbital frequency for the same binary configuration as the one for Fig.\ref{fig:freqs1}, but now with a mass ratio $q = M_1/M_2 = 3/2$. 

\begin{figure}
\includegraphics[width=3.25in]{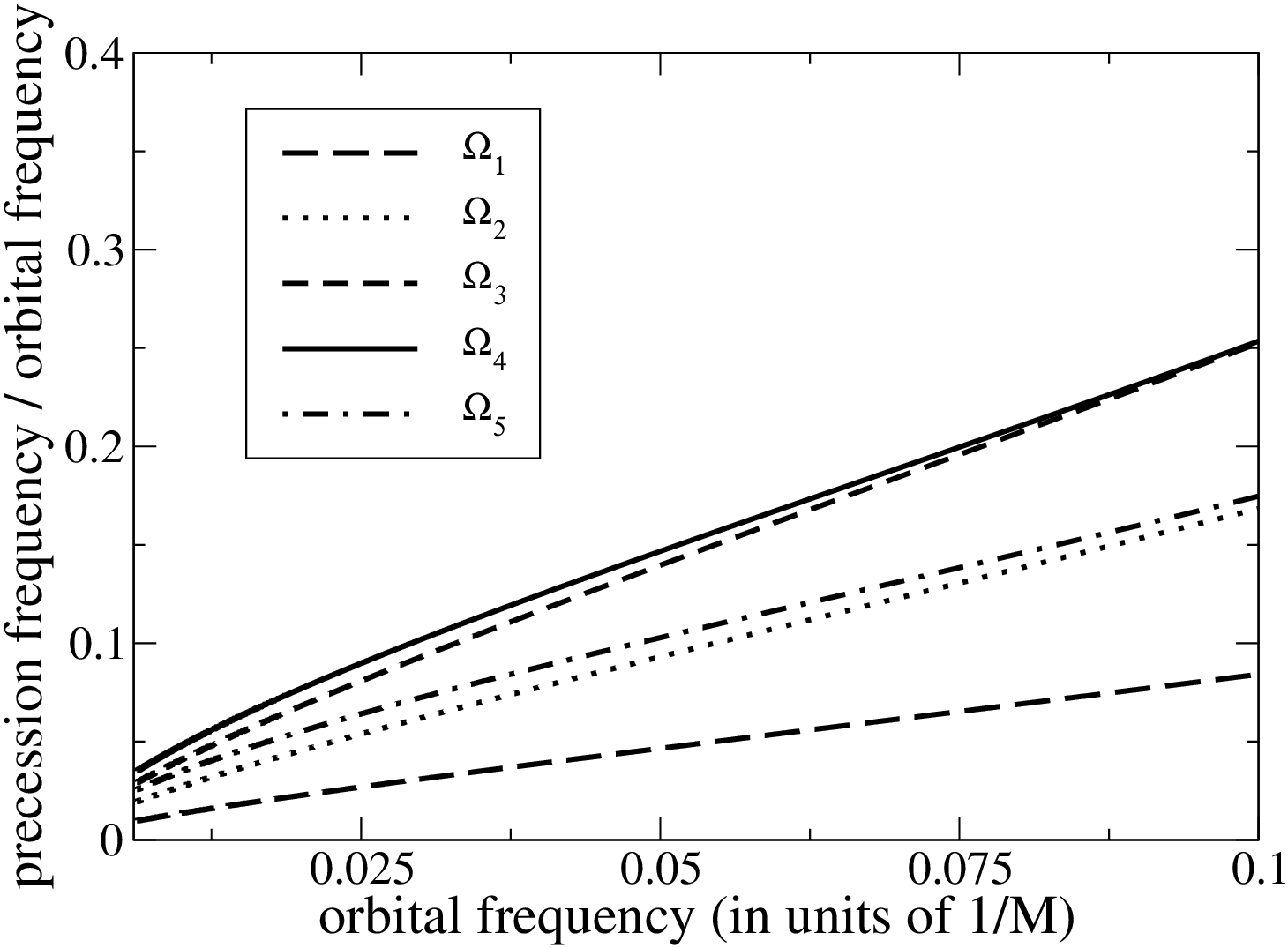}
\caption{\label{fig:freqs2} This is a plot of the precession frequencies appearing in our approximate spins, a mass ratio $q=M_1/M_2 = 3/2$, as function of orbital frequency. The binary configuration at orbital frequency of 20 Hz is specified by the following parameters: $M = 10\, M_\odot$, $\chi_1 = \chi_2 = 0.8$, $\theta_1 = \theta_2 = 30^\circ$, $\phi_1 = 0^\circ$ and $\phi_2 = 90^\circ$. The precession frequencies are expressed in units of instantaneous orbital frequency. The frequencies shown in the legend are defined as $\Omega_1 = d\Phi / dt$, $\Omega_2 = 2d\Phi / dt$, $\Omega_3 = 3d\Phi / dt$ , $\Omega_4 = d\Phi_+/dt$ and $\Omega_5 = d\Phi_-/dt$. Note however that the true frequency spectrum contains an infinite number of components, but these five are sufficient to obtain good accuracy.}
\end{figure}

This completes our construction of approximate solutions for the spins in presence of radiation reaction. We conclude this paper's analysis by comparing our analytic solutions to results obtained via numerical integration of the precession equations, including radiation reaction.

\subsection{Comparison with numerical solutions}\label{sec:comp}

\begin{figure}
\includegraphics[width=3.25in]{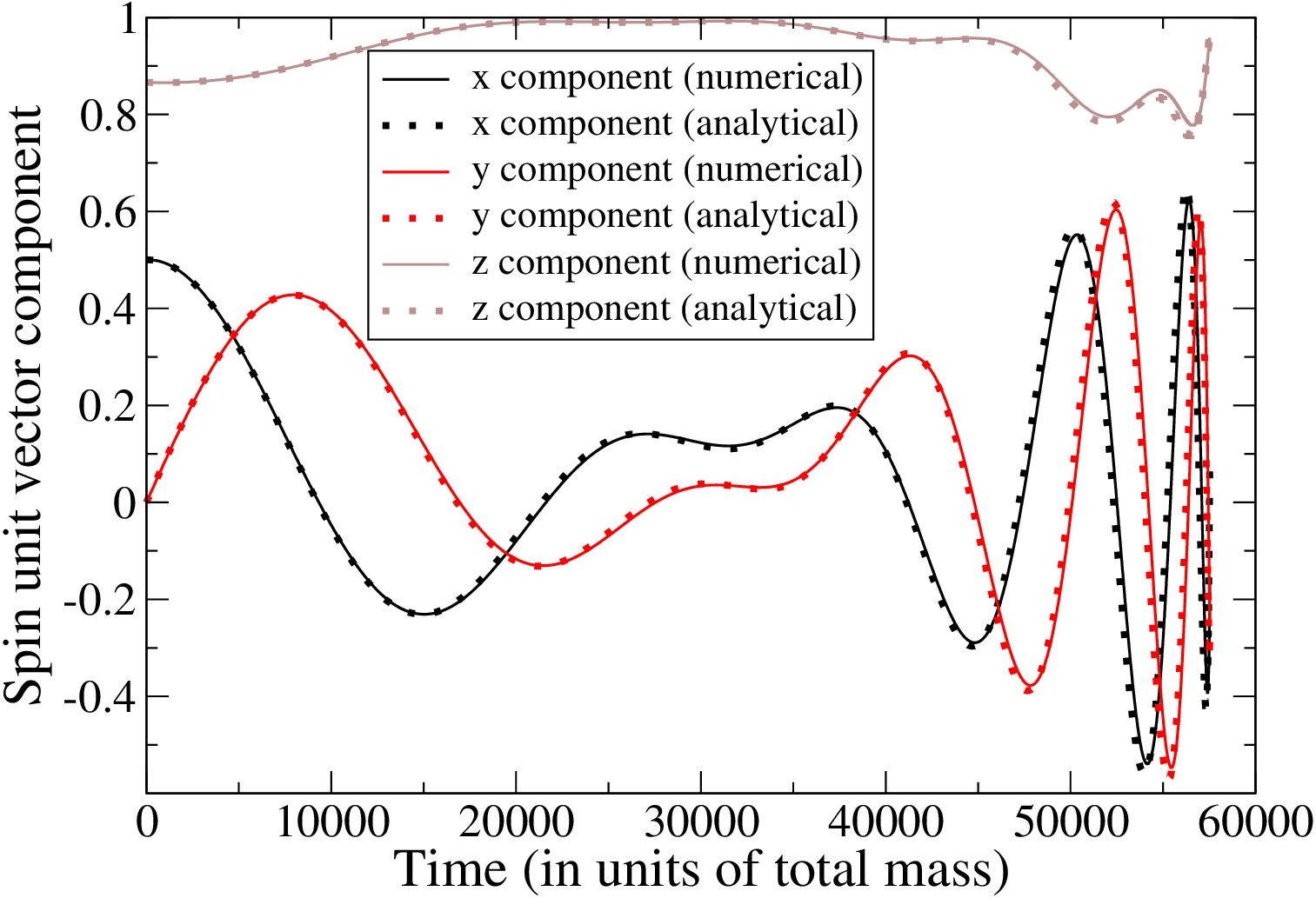}
\caption{\label{fig:spins1} This is a plot of the components of the unit vector $\hat{\bm{S}}_1$, for an equal mass binary. The binary configuration is the same as in Fig.\ref{fig:freqs1}. The $z$-axis is the initial orbital angular momentum axis, and the $x$-axis is along the initial direction of the component of $\bm{S}_1$ lying in the initial orbital plane.}
\end{figure}

\begin{figure}
\includegraphics[width=3.25in]{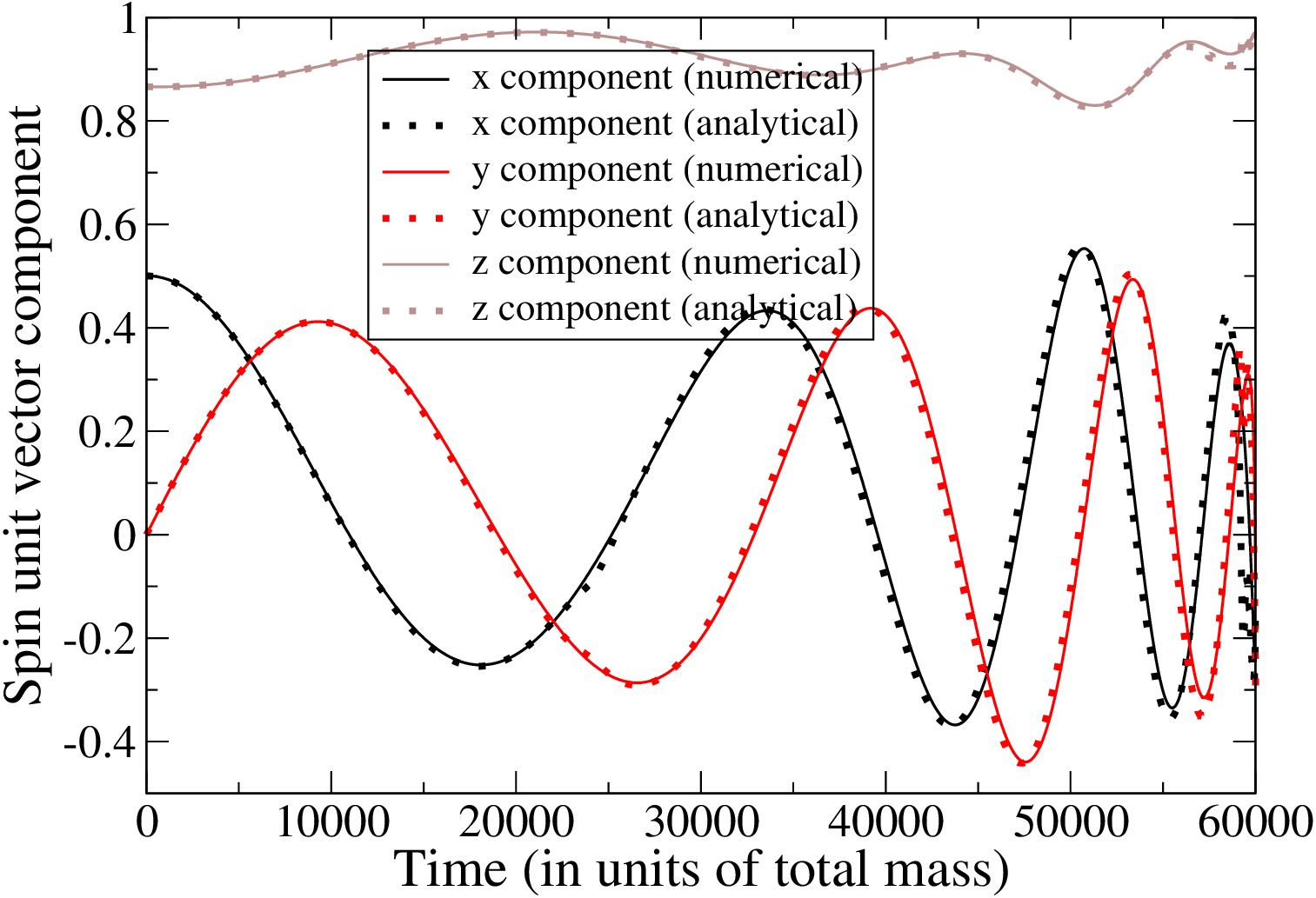}
\caption{\label{fig:spins2} This is a plot of the components of the unit vector along $\hat{\bm{S}}_1$, for a mass ratio $q = M_1/M_2 = 3/2$. The binary configuration is the same as in Fig.\ref{fig:freqs2}. Again the $z$-axis is the initial orbital angular momentum axis, and the $x$-axis is along the initial direction of the component of $\bm{S}_1$ lying in the initial orbital plane.}
\end{figure}

In this section we compare our approximate analytic solutions with solutions to the spin precession equations obtained numerically.  We begin by simply plotting the components of the unit vector along $\bm{S}_1$, for the same binary configurations used to generate figures \ref{fig:freqs1} and \ref{fig:freqs2}, to give a rough illustrative estimate of the accuracy of our approximate solutions. Figures \ref{fig:spins1} and \ref{fig:spins2} are generated as follows. First, we integrate the orbital frequency evolution equation coupled with the precession equations numerically to generate the function $\omega_{\rm num}(t)$. Since our spins are viewed as functions of frequency when including radiation-reaction [cf. computation of the phases in previous subsection], the quantities that are plotted in Figs. \ref{fig:spins1} and \ref{fig:spins2} are the components $\hat{\bm{S}}_{1,2}[\omega_{\rm num}(t)]$. Thus we here first verify the accuracy of the functional form of our spins.

Figures \ref{fig:spins1} and \ref{fig:spins2} both show excellent agreement between the numerical results and analytical predictions. Over most of the inspiral, the relative difference between the numerical and analytical solutions is less than a percent, except near the end of integration, which we set when $\omega_{\rm orb} = 0.1$. Also one can notice in Fig. \ref{fig:spins2} that around $t = 2.5 \times 10^4\, M$, the discrepancy between the analytical and numerical solutions gets a bit worse. This happens when the total spin precession goes through turning points, as pointed out in the previous discussion of Fig.\ref{fig:angles}. This effect is simply not captured as well as the rest of the evolution of the spins by our analytic solutions. One would need to incorporate more terms oscillating at different frequencies when solving (\ref{dotperpsII}) in order to improve our solutions in that regime. Near the end of integration, the errors in the spin components are generically in the range $\sim 10\% - 25\%$. 

We now move on to a different comparison, which is performed as follows. We first compute the orbital frequency as function of time by solving (\ref{omega3.5PN}) and the spin precession equations including radiation reaction simultaneously, the result being a function of time denoted as $\omega_{\rm num}(t)$, as before. One can then invert this equation and obtain time as function of frequency, i.e. $t = t(\omega_{\rm num})$.

We now turn to coupling our analytic solutions for the spins with (\ref{omega3.5PN}). To do this, we simply substitute the frequency dependent spins constructed in this section into the right-hand side of (\ref{omega3.5PN}), and then integrate the resulting differential equation with respect to time numerically, the result being a function of time denoted as $\omega_{\rm an}(t)$. We found generically that the coalescence time $t_{\rm an}$ where $\omega_{\rm an}$ diverges is shorter than the coalescence time associated with $\omega_{\rm num}$. This will be relevant below.

\begin{figure}
\includegraphics[width=3.25in]{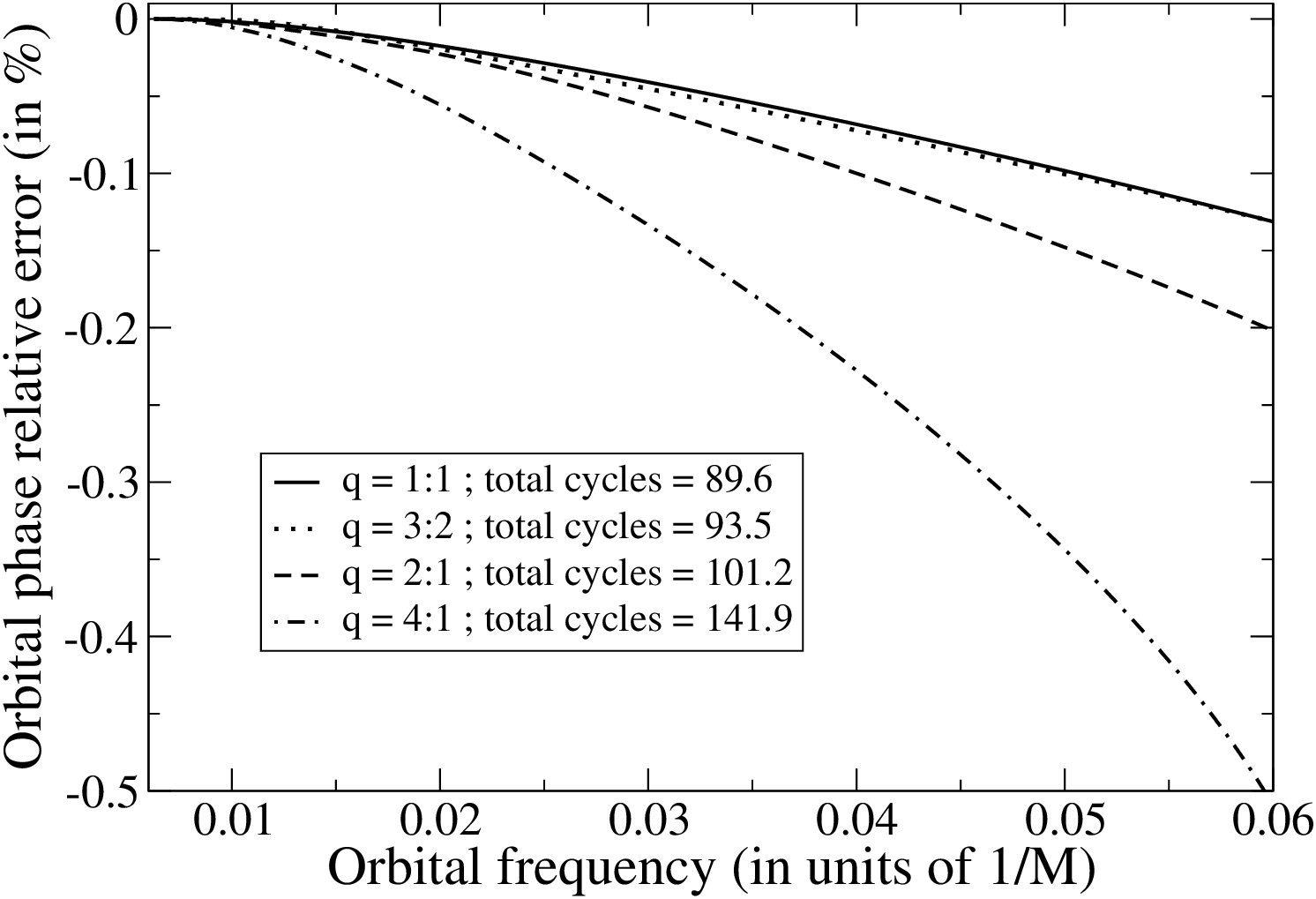}
\caption{\label{errorplot} This is a plot of accumulated orbital phase difference as function of orbital frequency for 4 different mass ratios (indicated in the legend), and the same binary parameters as in Fig.\ref{fig:freqs1}.} 
\end{figure}

To compare these two solutions, we look at the difference in accumulated orbital phase $\Delta \Phi$ as function of frequency, defined as

\be
\Delta \Phi(\omega_{\rm num}) = \int_0^{t(\omega_{\rm num})} \left[\omega_{\rm num}(t^\prime) - \omega_{\rm an}(t^\prime)\right]dt^\prime.
\ee
We use the orbital phase for our comparison instead of gravitational wave phase here, since a typical waveform for a spinning black hole binary contains many different frequencies, which depend on the binary configuration. Thus to make our comparisons as uniform and simple as possible, we restrict attention in this paper to the orbital phase only. In Fig.\ref{errorplot} we show a plot of $\Delta\Phi / \Phi_{\rm num}$ as function of $\omega_{\rm num}$, where $\Phi_{\rm num} = \int_0^t \omega_{\rm num}(t^\prime) \, dt^\prime$, for the binary configuration corresponding to Fig. \ref{fig:freqs1}, and for different mass ratios. The results for a few other points in parameter space are summarized in table \ref{table}, the data of Fig.\ref{errorplot} corresponding to the first block reported in the table. In Table \ref{table}, the quantity $\Delta\Phi/\Phi_{\rm num}$ is evaluated at the coalescence time $t_{\rm an}$. 

\begin{table}
\caption{\label{table} This table displays the results of our comparison for a representative sample of binary parameter space. All binary configurations have a total mass $M = 10\, M_\odot$ and the initial spins have $\chi_1 = \chi_2 = \chi$. The initial azimuthal angles are given by $\varphi_1 = 0^\circ$ and $\varphi_2 = 90^\circ$.}
\begin{ruledtabular}
\begin{tabular}{|c|c|c|c|}
$\begin{array}{c} {\rm binary} \\ {\rm parameters} \end{array}$ & mass ratio &$\begin{array}{c} {\rm number \,\, of} \\ {\rm \,\, orbital \,\, cycles} \end{array}$& $10^2\frac{\Delta \Phi}{\Phi_{\rm num}}$  \\
\hline
$\begin{array}{c} \chi = 0.8 \\ \theta_1 = 30^\circ \\ \theta_2 = 30^\circ \end{array}$ & $\begin{array}{c} 1:1 \\ 3:2 \\ 2:1 \\ 4:1 \end{array}$ & $\begin{array}{c} 89.6 \\ 93.5 \\ 101.2 \\ 141.9 \end{array}$& $\begin{array}{r} -0.42  \\ -0.30 \\ -0.41 \\ -0.67 \end{array}$\\
\hline
$\begin{array}{c} \chi = 0.8 \\ \theta_1 = 90^\circ \\ \theta_2 = 90^\circ \end{array} $ & $\begin{array}{c} 1:1 \\ 3:2 \\ 2:1 \\ 4:1 \end{array}$ & $\begin{array}{c} 84.7 \\ 88.2 \\ 95.3 \\ 133.1 \end{array}$& $\begin{array}{r} -0.85  \\ -0.85 \\ -1.38 \\ -1.58 \end{array}$\\
\hline
$\begin{array}{c} \chi = 0.8 \\ \theta_1 = 150^\circ \\ \theta_2 = 150^\circ \end{array}  $& $\begin{array}{c} 1:1 \\ 3:2 \\ 2:1 \\ 4:1 \end{array}$ & $\begin{array}{c} 79.7 \\ 83.1 \\ 89.8 \\ 125.1 \end{array}$& $\begin{array}{r} -0.39  \\ -0.46 \\ -0.54 \\ -0.89 \end{array}$\\
\hline
$\begin{array}{c} \chi = 1.0 \\ \theta_1 = 30^\circ \\ \theta_2 = 30^\circ \end{array} $ & $\begin{array}{c} 1:1 \\ 3:2 \\ 2:1 \\ 4:1 \end{array}$ & $\begin{array}{c} 90.9 \\ 94.9 \\ 102.8 \\ 144.3 \end{array}$& $\begin{array}{r} -0.53  \\ -0.38 \\ -0.52 \\ -0.86 \end{array}$\\
\hline
$\begin{array}{c} \chi = 1.0 \\ \theta_1 = 90^\circ \\ \theta_2 = 90^\circ \end{array} $& $\begin{array}{c} 1:1 \\ 3:2 \\ 2:1 \\ 4:1 \end{array}$ & $\begin{array}{c} 84.7 \\ 88.2 \\ 95.4 \\ 133.1 \end{array}$& $\begin{array}{r} -1.15  \\ -1.08 \\ -1.35 \\ -1.84 \end{array}$\\
\hline
$\begin{array}{c} \chi = 1.0 \\ \theta_1 = 150^\circ \\ \theta_2 = 150^\circ \end{array} $& $\begin{array}{c} 1:1 \\ 3:2 \\ 2:1 \\ 4:1 \end{array}$ & $\begin{array}{c} 78.5 \\ 81.8 \\ 88.5 \\ 123.2 \end{array}$& $\begin{array}{r} -0.55  \\ -0.62 \\ -0.76 \\ -0.95 \end{array}$\\
\end{tabular}
\end{ruledtabular}
\end{table}

Generically, the errors are larger for bigger mass ratios since it takes more orbital cycles to sweep through a given frequency band. Thus there is more time for the errors inherent to our approximation scheme to build up. We believe this simple argument explains the generic features of Fig.\ref{errorplot}, and Table \ref{table}. The relative phase errors range from about 0.3\% to  about 1.8\% for $\sim 80 - 140$ orbital cycles. This corresponds to errors of order $\sim 6$ radians over $\sim 100$ orbital cycles, which is a little too large for direct implementation in template banks, which require $\sim 1$ radian of accuracy over $\sim 500$ orbital cycles. Thus our solutions would need additional refinement before being suitable for template bank generation. However we believe that this $\sim 1\%$ accuracy is quite sufficient for a first cut at comparing the results of numerical relativity with post-Newtonian theory.

\section{Conclusion}

We presented a detailed analysis of the spin precession equations in black hole binaries, when the quadrupole-monopole contribution is taken into account. We showed that the (orbit-averaged) precession equations supplemented by the quadrupole-monopole term possess a conserved quantity, which had not been previously noticed. The existence of this conserved quantity allowed us to solve the precession equations exactly for arbitrary spins in the equal mass case, neglecting radiation reaction. When masses are unequal, we resorted to a perturbative expansion to solve the precession equations approximately, as an exact solution could not be obtained in that case. We then showed how to incorporate radiation reaction effects into our solutions for the spins adiabatically. To assess the accuracy of our approximate analytic solutions in the context of gravitational wave data analysis, we integrated the evolution equation for the orbital frequency accurate to 3.5PN order using our analytic spins, and also using the solution for the spins obtained from numerically integrating the precession equations. We showed that the relative errors in the accumulated orbital phase is of order $\sim 0.3\% - 1.8\%$ over $\sim 80 - 140$ orbital cycles, the errors being the worst when the spins are initially lying in the orbital plane. Our solutions however turn out to be quite complicated, which will probably make them unattractive for direct implementation in template families. Nevertheless we believe that they are potentially quite useful for phenomenological analysis of numerical models of spinning binary black holes, e.g. those provided by the numerical relativity community, and to improve general understanding of precession phenomena in binary systems.

\begin{acknowledgements}
The author wishes to thank Alessandra Buonanno for many helpful discussions and invaluable advice during the completion of this work. This research was partially supported by NSF grant PHY-0603762.
\end{acknowledgements}


\begin{thebibliography}{0}

\bibitem{LIGO}
A. Abramovici {\it et al.}, Science {\bf 256}, 325 (1992).

\bibitem{VIRGO}
B. Caron {\it et al.}, Class. Quant. Grav. {\bf 14}, 1461 (1997).

\bibitem{GEO}
H. L\:{u}ck {\it et al.}, Class. Quant. Grav. {\bf 14}, 1471 (1997).

\bibitem{TAMA}
M. Ando {\it et al.}, Phys. Rev. Lett. {\bf 86}, 3950 (2001).

\bibitem{BlanchetLiving}
L. Blanchet, {\it Gravitational Radiation from Post-Newtonian Sources and Inspiralling Compact Binaries},
Living Rev. Relativity {\bf 9},  (2006).

\bibitem{Apostolatos}
T. A. Apostolatos, C. Cutler, G. J. Sussman and K. S. Thorne, Phys Rev. D {\bf 49}, 6274 (1994).

\bibitem{BCV2}
A. Buonanno, Y. Chen and M. Vallisneri, Phys. Rev. D {\bf 67}, 104025 (2003); Erratum-ibid. {\bf 74} 029904 (2006).

\bibitem{Pan}
Y. Pan, A. Buonanno, Y. Chen and M. Vallisneri, Phys. Rev. D {\bf 69}, 104017 (2004); Erratum-ibid. {\bf 74} 029905 (2006).

\bibitem{Kidder}
L. E. Kidder, Phys. Rev. D {\bf 52}, 821 (1995).

\bibitem{BBF}
G. Faye, L. Blanchet and A. Buonanno, Phys. Rev. D {\bf 74}, 104033 (2006).

\bibitem{barker}
B. M. Barker and R. F. O'Connell, Gen. Relativ. Gravit. {\bf 11}, 149 (1979).

\bibitem{EOB}
T. Damour, Phys. Rev. D {\bf 64}, 124013 (2001).

\bibitem{Poisson}
\'{E}. Poisson, Phys. Rev. D {\bf 57}, 5287 (1998).

\bibitem{g1}
L. \'{A}. Gergely and  Z. Keresztes, Phys. Rev. D {\bf 67}, 024020 (2003).

\bibitem{g2}
Z. Keresztes, B. Mik\'{o}czi and  L. \'{A}. Gergely, Phys. Rev. D {\bf 72}, 104022 (2005).

\bibitem{DSX}
T. Damour, M. Soffel and C. Xu, Phys. Rev. D {\bf 45}, 1017 (1992); T. Damour, M. Soffel and C. Xu, Phys. Rev. D {\bf 47}, 3124 (1993).

\bibitem{Racine}
\'{E}. Racine, Class. Quant. Grav. {\bf 23},  373 (2006).

\bibitem{ThorneHartle}
K. S. Thorne and J. B. Hartle, Phys. Rev. D {\bf 31}, 1815 (1985).

\bibitem{JS}
J. D. Schnittman, Phys. Rev. D {\bf 70}, 124020 (2004).

%\bibitem{BFIJ}
%L. Blanchet, G. Faye, B. R. Iyer and B. Joguet, Phys. Rev. D {\bf 65}, 061501 (2002); Erratum-ibid. {\bf 71} 129902 (2005);

\bibitem{BBOC}
B. M. Barker, G. G. Byrd and R. F. O'Connell, Ap. J. {\bf 253}, 309 (1982).

\bibitem{WW}
E. T. Whittaker and G. N. Watson, {\it A Course in Modern Analysis (4th ed.)}, Cambridge University Press, New York, NY, (1927).

\bibitem{Goldstein}
H. Goldstein, {\it Classical Mechanics (2nd ed.)}, Addison-Wesley, Reading, MA, (1980). 

\bibitem{Mikoczi}
B. Mik\'{o}czi, M. Vas\'{u}th and L. \'{A}. Gergely, Phys. Rev. D {\bf 71}, 124043 (2005).

\end{thebibliography}
\end{document}